\documentclass[twocolumn,twocolappendix,appendixfloats]{openjournal}
\usepackage{graphicx}
\usepackage{amsmath}   % must come before txfonts, which redefines \iint etc.
\usepackage{txfonts}
\usepackage[]{hyperref}
\hypersetup{
   colorlinks,%
   citecolor=blue,%
   filecolor=blue,%
   linkcolor=blue,%
   urlcolor=blue
}
\usepackage{bm}
\providecommand{\mathds}[1]{\mathbf{#1}}
\usepackage{float}
\usepackage{subcaption}
\usepackage{array}
\renewcommand{\arraystretch}{1.2} % Augmente l'espacement des lignes par un facteur de 1.2
\usepackage{orcidlink}

\usepackage{ulem}
\newcommand{\unit}[1]{
\mathrm{#1}
}

\newcommand{\averagePhi}[1]{
\left\langle #1 \right\rangle
}

\newcommand{\mH}{
m_{\mathrm{H}}
}

\newcommand{\kB}{
k_{\mathrm{B}}
}

\makeatletter
\expandafter\let\csname longtable*\endcsname\@undefined
\expandafter\let\csname endlongtable*\endcsname\@undefined
\makeatother

\begin{document} 

\title{Angular momentum transport in a newborn protostar--disc system: magnetic braking, spiral arms and the magneto-rotational instability}

\author{Antonin Borderies \orcidlink{0009-0005-5957-9429}}
\email{antonin.borderies@yahoo.fr}
\author{Benoît Commerçon \orcidlink{0000-0003-2407-1025}}
\author{Adnan Ali Ahmad \orcidlink{0009-0004-2719-1107}}
\affiliation{ENS de Lyon, Université Lyon 1, CNRS, CRAL, UMR 5574, Lyon, France}

\begin{abstract}
    Angular momentum transport in young protostar--disc systems involves magnetic braking,
    magneto-rotational turbulence and gravitational instability, mechanisms that are usually
    studied separately and on different scales. We present a three-dimensional non-ideal
    magnetohydrodynamic simulation, including Ohmic and ambipolar diffusion, that follows the
    collapse of a $1~M_\odot$ dense core continuously from the cloud scale to the formation of the
    protostar, and then the resulting protostar--disc system for $\sim 64~\mathrm{yr}$ at a
    resolution of $\sim 10^{-4}~\mathrm{AU}$ at the protostellar surface. We decompose the
    azimuthally averaged angular momentum budget into Reynolds, Maxwell and gravitational torques,
    each split into a radial and a meridional component, over a quasi-steady phase of the
    evolution. The collapse produces a thick disc an order of magnitude more massive than the
    protostar, whose specific angular momentum decreases as it accretes. Accretion through the disc
    is vertically layered, with accreting and decreting layers at the same radius, a pattern set
    layer by layer by the torques. The transport separates into three regimes delimited by the
    coupling of the magnetic field to the gas. Magnetic braking operates in the envelope. Maxwell
    stresses dominate inside $\sim 0.1~\mathrm{AU}$, where the field recouples and the
    magneto-rotational instability develops. In the ambipolar dead zone that makes up the bulk of
    the disc, the Reynolds stress dominates, carried by $m=2$ spiral arms in a marginally unstable
    disc, with the gravitational stress as a secondary contributor. Meridional transport is as
    important as radial transport, so that a one-dimensional viscous model fitted to the net mass
    flux would require a negative $\alpha$ over part of the disc.
\end{abstract}

% Write your keywords here

\maketitle

%
%-------------------------------------------------------------------

%%%%%%%%%%%%%%%%%%%%%%%%%%%%%%%%%%%%%%%%%%%%%%%%%%%%%%%%%%%%%%%%%%%%%%%%%%%%%%%%%%%%%%%%%%%%
%%%%%%%%%%%%%%%%%%%%%%%%%%%%%%%%%%%%%%%%%%%%%%%%%%%%%%%%%%%%%%%%%%%%%%%%%%%%%%%%%%%%%%%%%%%%
% \nolinenumbers{}
\section{Introduction}
\label{sec: introduction}

Star formation is in large part a problem of angular momentum disposal. A dense core inherits
rotation from its parent cloud, and as it contracts it spins up, so that by the time material
reaches the scale of the forming star it carries far more specific angular momentum than the
star itself can hold. The gas therefore cannot fall directly to the centre: deflected by
rotation, it settles into a rotationally supported circumstellar disc. The disc, however, only
defers the difficulty. For the protostar to grow, the material orbiting it must give its
angular momentum to something else --- accretion is less a matter of drawing gas inwards than
of removing the quantity that keeps it out. Whatever performs that removal sets the rate at
which the star is assembled, and the mass, size and structure of the disc through which it is
assembled. Such discs are now observed around Class~0 protostars
\citep{tobin02solarmassProtostarKeplerian2012,ohashiFormationKeplerianDisk2014}, and surveys
find most of them to be compact, with radii of a few tens of astronomical units
\citep{mauryCharacterizingYoungProtostellar2019,tobinVLAALMANascent2020}. The protostar itself,
however, is only a few solar radii across, so the angular momentum must ultimately be removed
down to scales of $\sim 0.01~\unit{AU}$, well below what these observations resolve.

Three mechanisms are generally invoked to remove it. In the magnetised envelope, the field
couples the rapidly rotating inner regions to the slowly rotating outer ones and brakes them
\citep{mouschoviasMagneticBrakingAligned1980,hennebelleMagneticProcessesCollapsing2008}. Under
ideal magnetohydrodynamics this braking is so efficient that no rotationally supported disc
survives the collapse \citep{mellonMagneticBrakingProtostellar2008}; discs form because Ohmic
and ambipolar diffusion decouple the field from the weakly ionised gas
\citep{marchandChemicalSolverCompute2016,zhaoProtostellarDiscFormation2016,tsukamotoRoleMagneticFields2023},
leaving extended dead zones in which magnetic stresses cannot operate. Where the field remains
well coupled to a disc, the magneto-rotational instability (MRI) provides turbulent transport
\citep{balbusPowerfulLocalShear1991,hawleyLocalThreedimensionalMagnetohydrodynamic1995}.
Finally, in the massive discs that emerge from a collapse, gravitational instability drives
spiral arms that redistribute angular momentum
\citep{toomreGravitationalStabilityDisk1964,kratterGravitationalInstabilitiesCircumstellar2016}.

These three are usually studied apart from one another. The division is visible in the
structure of the field itself: the lectures collected from the school \emph{Role and mechanisms
of angular momentum transport in the formation and early evolution of stars} treat angular
momentum in collapsing dense cores \citep{hennebelleTransportInfluenceAngular2013} and in
protoplanetary discs \citep{fromangMRIdrivenAngularMomentum2013} separately, and largely from
separate literatures. The reason is methodological. Collapse calculations resolve the envelope
and the forming disc, but either stop at the first hydrostatic core or, when they reach the
protostar, follow it only briefly. Disc calculations begin instead from an already-formed disc
of prescribed structure, without the collapse history that fixes its mass and its magnetic
field. Most simulations that follow the collapse down to the
protostar
\citep{vaytetProtostellarBirthAmbipolar2018,bhandareBirthConvectiveLowmass2020,ahmadBirthEarlyEvolution2023,ahmadBirthMagnetizedLowmass2025}
integrate for months to a few years after its formation. A notable exception is
\citet{machidaFirstTwoThousand2019}, who followed the system for $\sim 2000~\unit{yr}$ and
found a disc in which, once the field has dissipated, gravitational torques dominate the
transport and drive episodic accretion onto the protostar. Their calculation, however, included
Ohmic dissipation only, and a resolution of $5.6\times10^{-3}~\unit{AU}$ at the protostar,
more than an order of magnitude coarser than here. The relative contribution of the different mechanisms has therefore
not yet been measured from the full angular momentum budget of a system in which ambipolar
diffusion is included and the protostar--disc interface is resolved.

This is what the present paper does. Using the finite-volume code \textsc{Idefix}
\citep{lesurIDEFIXVersatilePerformanceportable2023}, we follow the collapse of a dense core
with Ohmic and ambipolar diffusion continuously from the cloud scale to the formation of the
protostar, with a relative resolution $\Delta r/r \simeq 2\%$ that gives cells of
$\sim 10^{-4}~\unit{AU}$ at the protostellar surface, and then integrate the resulting
protostar--disc system for some tens of years, long enough for the inner disc to reach a
quasi-steady state. We then analyse the angular momentum budget of that system directly,
decomposing it into the Reynolds, Maxwell and gravitational stresses and splitting each into a
radial and a meridional component. The decomposition is carried out in spherical geometry, and
deliberately retains the vertical direction: the disc that emerges from the collapse is
geometrically thick and an order of magnitude more massive than the protostar it surrounds, so
neither the thin-disc approximation nor the assumption of a dominant central mass applies, and
the vertical transport turns out to be as important as the radial transport in setting where
material goes.

The paper is organised as follows. Section~\ref{sec: method} describes the physical model, the
numerical setup and the restart strategy that makes a single calculation spanning seven decades
in radius possible, together with the diagnostics used to quantify the transport.
Section~\ref{sec: overview} presents the resulting protostar--disc system,
Sect.~\ref{sec: angular_momentum_equilibrium} establishes that its accretion pattern is
controlled by angular momentum transport, and Sect.~\ref{sec: transport_mechanisms} identifies
which stress dominates in which region, expressing the result as an equivalent $\alpha$
parameter. Section~\ref{sec: origin} examines the origin of the magnetic and gravitational
stresses and of the mean vertical motion in the disc, Sect.~\ref{sec: discussion} discusses the
limitations of the models and summarises our conclusions.

%%%%%%%%%%%%%%%%%%%%%%%%%%%%%%%%%%%%%%%%%%%%%%%%%%%%%%%%%%%%%%%%%%%%%%%%%%%%%%%%%%%%%%%%%%%%
%%%%%%%%%%%%%%%%%%%%%%%%%%%%%%%%%%%%%%%%%%%%%%%%%%%%%%%%%%%%%%%%%%%%%%%%%%%%%%%%%%%%%%%%%%%%
\section{Method}
\label{sec: method}

Our goal is to follow the gravitational collapse of a dense core continuously, from the
initial cloud down to the formation of the protostar, and then to integrate the resulting
protostar--disc system for as long as possible. Simulating the whole collapse provides
self-consistent initial conditions for the protostar--disc system, at the price of covering
seven orders of magnitude in spatial scale within a single run. In this section we describe
the physical model (Sect.~\ref{subsec: physical_model}), the initial conditions
(Sect.~\ref{subsec: initial_conditions}) and the grid (Sect.~\ref{subsec: grid}), the restart
strategy used to follow the successive scales of the collapse
(Sect.~\ref{subsec: restart}), the boundary conditions (Sect.~\ref{subsec: boundary}), the
internal limiters required to keep the integration tractable
(Sect.~\ref{subsec: limiters}), and finally the diagnostics used to quantify angular
momentum transport (Sect.~\ref{subsec: angular_momentum_budget}).

%-------------------------------------------------------------------
\subsection{Physical model}
\label{subsec: physical_model}

\subsubsection{Equations solved}

We solve the non-ideal magnetohydrodynamics (MHD) equations with self-gravity,
including Ohmic and ambipolar diffusion:
\begin{equation}
    \begin{cases}
        \partial_t \rho + \nabla \cdot (\rho \bm{v}) = 0,\\[3pt]
        \partial_t (\rho \bm{v}) + \nabla \cdot \left( \rho \bm{v} \otimes \bm{v}
        - \dfrac{\bm{B} \otimes \bm{B}}{4\pi} + P_\mathrm{tot}\mathds{1} \right)
        = -\rho \nabla \Phi,\\[3pt]
        \partial_t \bm{B} = \nabla \times \left(\bm{v} \times \bm{B}
        - \eta_\Omega \bm{J} + \eta_\mathrm{A} \left(\bm{J} \times \bm{b}\right) \times \bm{b}\right),\\[3pt]
        \nabla^2 \Phi = 4\pi G \rho,
    \end{cases}
    \label{eq: mhd_system}
\end{equation}
where $\rho$ is the gas density, $\bm{v}$ the velocity, $\bm{B}$ the magnetic field,
$\Phi$ the gravitational potential and $\mathds{1}$ the identity tensor. We have written
$P_\mathrm{tot} = P(\rho) + B^2/8\pi$ for the total (thermal plus magnetic) pressure,
$\bm{J} = \nabla \times \bm{B}$, and $\bm{b} = \bm{B}/B$. The quantities $\eta_\Omega$ and
$\eta_\mathrm{A}$ are the Ohmic and ambipolar diffusivities. No energy equation is solved,
since the thermodynamics is closed by a barotropic equation of state $P(\rho)$ (Sect.~\ref{subsubsec: eos}). No explicit viscosity
is included: the molecular viscosity of the gas is entirely negligible under star-formation
conditions, and we deliberately introduce no artificial shear viscosity, so that the angular
momentum transport analysed in this paper is produced by the flow itself.

\subsubsection{Equation of state}
\label{subsubsec: eos}

\begin{figure}[t]
    \centering
    \includegraphics[width=0.8\columnwidth]{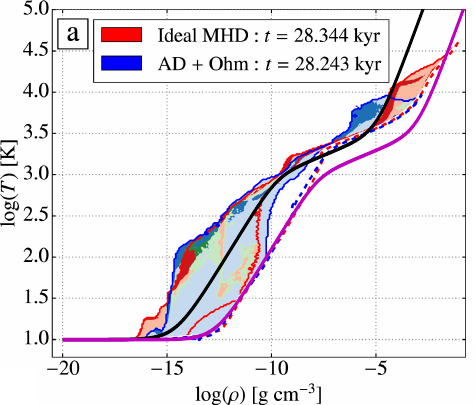}
    \caption{Temperature--density relation adopted in this work. The dashed lines show the
    evolution of the temperature at the centre of an MHD collapse simulation with radiative
    transfer, and the shaded areas the final state of the medium after second core formation,
    both from \citet{vaytetProtostellarBirthAmbipolar2018}. The purple solid line is the
    barotropic law of Eq.~\eqref{eq: barotropic_law} with the critical densities of
    \citet{marchandChemicalSolverCompute2016}; the black solid line is the same law with the
    critical densities reduced by a factor of 100, which is the prescription used here.}
    \label{fig: temperature_baro}
\end{figure}

We close the system with a barotropic equation of state, $P = \rho \kB T(\rho) / (\mu_\mathrm{mol}\mH)$,
with $\mu_\mathrm{mol} = 2.34$, and with the temperature--density relation of
\citet{marchandChemicalSolverCompute2016},
\begin{equation}
    T(\rho) = T_0\sqrt{1+\left(\frac{\rho}{\rho_1}\right)^{2g_1}}
    \left(1+\frac{\rho}{\rho_2}\right)^{g_2}\left(1+\frac{\rho}{\rho_3}\right)^{g_3},
    \label{eq: barotropic_law}
\end{equation}
with $T_0 = 10~\unit{K}$, $g_1 = 0.4$, $g_2 = -0.3$ and $g_3 = 0.56667$. This functional form
reproduces the four successive phases of the collapse: the isothermal phase
($\rho \ll \rho_1$), the first adiabatic core ($\rho_1 \ll \rho \ll \rho_2$, effective
$\gamma_\mathrm{eff} = 7/5$), the second collapse driven by $\mathrm{H_2}$ dissociation
($\rho_2 \ll \rho \ll \rho_3$, $\gamma_\mathrm{eff} = 1.1$), and the second Larson core
($\rho_3 \ll \rho$, $\gamma_\mathrm{eff} = 5/3$).

The critical densities of \citet{marchandChemicalSolverCompute2016}
($\rho_1 = 3.9\times 10^{-13}$, $\rho_2 = 3.9\times 10^{-8}$ and
$\rho_3 = 3.9\times 10^{-3}~\unit{g\,cm^{-3}}$) accurately reproduce the central temperature
during the first collapse, but underestimate it during the second collapse, and they always
underestimate the temperature in the outer layers of the protostar, where accretion shocks and
radiative transport deposit energy (Fig.~\ref{fig: temperature_baro}). Using them yields a
protostar smaller than $0.1~R_\odot$, which would be prohibitively expensive to resolve. We
therefore keep the functional form of Eq.~\eqref{eq: barotropic_law} but reduce the three
critical densities by a factor of 100, i.e. $\rho_1 = 3.9\times 10^{-15}$,
$\rho_2 = 3.9\times 10^{-10}$ and $\rho_3 = 3.9\times 10^{-5}~\unit{g\,cm^{-3}}$, which yields
a protostar of $\sim 1~R_\odot$. Although the rescaled law no longer reproduces the evolution of
the central temperature, it gives a more faithful description of the temperature of the
surrounding gas after second core formation (shaded areas in Fig.~\ref{fig: temperature_baro}).
This rescaling has no thermodynamical justification beyond
matching the protostellar radius and the general temperature--density behaviour found in
radiation-hydrodynamics calculations; our simulations are consequently not predictive
regarding the thermal evolution of the system.

\subsubsection{Non-ideal MHD resistivities}
\label{subsubsec: resistivities}

The Ohmic and ambipolar diffusivities are read from the pre-computed tables of
\citet{marchandChemicalSolverCompute2016}, as functions of the density, the temperature and
the magnetic field strength, assuming an MRN grain size distribution. This standard approach
assumes that the microphysics of the gas--dust mixture relaxes on a timescale much shorter
than the dynamical timescale, so that the ionisation state is always in equilibrium with the
local conditions. This assumption is questionable in the innermost regions, in particular
regarding dust evolution \citep{borderiesDustEvolutionChemisputtering2025}, and we return to
its consequences in Sect.~\ref{sec: discussion}.

\subsubsection{Numerical scheme}
\label{subsubsec: numerical_scheme}

The system~\eqref{eq: mhd_system} is integrated with the finite-volume code \textsc{Idefix}
\citep{lesurIDEFIXVersatilePerformanceportable2023}, which is built on the \textsc{Kokkos}
performance-portability library \citep{trottKokkos3Programming2022} and runs on GPUs. We use
the HLLD approximate Riemann solver \citep{miyoshiMultistateHLLApproximate2005}, second-order
piecewise-linear reconstruction with the van Leer slope limiter
\citep{vanleerUltimateConservativeDifference1979}, and a second-order Runge--Kutta time
integrator. The choice of Riemann solver is not innocuous for the problem at hand: the more
diffusive HLL solver \citep{hartenUpstreamDifferencingGodunovType1983,
einfeldtGodunovTypeMethodsLow1991} suppresses the magneto-rotational instability at the
resolutions we can afford \citep[see also][]{flockHighorderGodunovSchemes2010}. We
therefore adopt HLLD throughout the main text, and report the corresponding HLL run in
Appendix~\ref{app: hll}.

The solenoidality of the magnetic field is enforced by constrained transport
\citep{evansSimulationMagnetohydrodynamicFlows1988}, in which we evolve the vector potential
$\bm{A}$ at the cell edges rather than the face-centred field itself, and recover $\bm{B}$ from
its circulation. The Poisson equation is solved
with a BiCGSTAB iterative solver \citep{saadIterativeMethodsSparse2003}, as implemented in
\textsc{Idefix} by \citet{mauxionInteractionDisquesProtoplanetaires2024}, and the
gravitational acceleration is obtained from a fourth-order finite-difference estimate of
$\nabla\Phi$. The parabolic (diffusive) terms are treated explicitly with a second-order
Runge--Kutta--Legendre super-time-stepping scheme
\citep{meyerStabilizedRungeKutta2014,vaidyaScalableExplicitImplementation2017}, which covers a
hyperbolic time step $\Delta t_\mathrm{hyp}$ in $\sim 2\sqrt{\zeta}$ sub-steps instead of
$\sim 2\zeta$, where $\zeta = \Delta t_\mathrm{hyp}/\Delta t_\mathrm{par}$ is the ratio of the
hyperbolic to the parabolic time step. Even so, $\zeta$ can exceed $10^6$ in a protostellar
collapse, which is why an additional diffusivity limiter is required
(Sect.~\ref{subsubsec: diffusivity_limiter}).

%-------------------------------------------------------------------
\subsection{Initial conditions}
\label{subsec: initial_conditions}

We adopt initial conditions similar to those of
\citet{vaytetProtostellarBirthAmbipolar2018}: a uniform-density sphere of mass $M_0$ and
radius $R_0$, in solid-body rotation at angular velocity $\Omega_0$, threaded by a uniform
magnetic field $B_0$ aligned with the rotation axis. For a given $M_0$, the three parameters
$R_0$, $\Omega_0$ and $B_0$ are set by the dimensionless ratios of thermal, rotational and
magnetic energy to the gravitational energy of the cloud,
\begin{equation}
    \alpha = \frac{5}{2}\frac{\kB T_0 R_0}{\mu_\mathrm{mol} \mH G M_0}, \quad
    \beta = \frac{1}{3}\frac{\Omega_0^2 R_0^3}{G M_0}, \quad
    \mu = \frac{M_0}{\pi R_0^2 B_0 \left(M/\Phi_B\right)_\mathrm{crit}},
\end{equation}
where $T_0 = 10~\unit{K}$ is the initial cloud temperature and
$\left(M/\Phi_B\right)_\mathrm{crit} = \frac{0.53}{3\pi}\left(5/G\right)^{1/2}$ is the critical
mass-to-flux ratio \citep{mouschoviasNoteCollapseMagnetic1976}. We adopt $M_0 = 1~M_\odot$,
$\alpha = 0.4$, $\beta = 0.01$ and $\mu = 4$, which
correspond to a gravitationally unstable, moderately magnetised and slowly rotating core.

The corresponding initial fields are
\begin{equation}
    \begin{cases}
        \rho = \rho_0 = \dfrac{3M_0}{4\pi R_0^3} \quad \text{if } r < R_0, \quad
        \rho_0/100 \ \text{ otherwise},\\[4pt]
        v_r = v_\theta = 0,\\[2pt]
        v_\varphi = \Omega_0 r \sin\theta \quad \text{if } r < R_0, \quad 0 \ \text{ otherwise},\\[4pt]
        A_\varphi = \dfrac{B_0}{2} r\sin\theta \quad \text{if } r\sin\theta < R_0,\\[4pt]
        \phantom{A_\varphi =} \dfrac{B_0}{2}\left(\dfrac{r\sin\theta - R_0}{100^{2/3}} + R_0 \right)
        \ \text{ otherwise}.
    \end{cases}
\end{equation}
The ambient medium outside the cloud is set to $\rho_0/100$ to avoid excessively low
densities, and the vector potential is scaled down outside the cloud following the
$B \propto \rho^{2/3}$ relation expected for spherical collapse. No azimuthal perturbation is added to the system
at this stage.

%-------------------------------------------------------------------
\subsection{Grid}
\label{subsec: grid}

\begin{table}[t]
    \centering
    \caption{Final grid extent and resolution, after all restarts.}
    \begin{tabular}{c|c|c|c}
        \hline\hline
        Direction & Lower & Upper & Resolution \\
        \hline
        $r$ & $10^{-3}~\unit{AU}$ & $10^{4}~\unit{AU}$ & 128 cells per decade \\
        $\theta$ & $0$ & $\pi$ & 128 cells \\
        $\varphi$ & $0$ & $\pi$ & 64 cells \\
        \hline
    \end{tabular}
    \label{tab: grid}
\end{table}

\textsc{Idefix} does not implement adaptive mesh refinement, so the resolution must be
concentrated on the region of interest by the choice of geometry. We use a spherical grid
$(r,\theta,\varphi)$ with logarithmic spacing in the radial direction, which gives a constant
relative resolution $\Delta r / r$ across the seven decades covered by the collapse. The cell
interfaces are placed at
\begin{equation}
    r_{i-1/2} = r_\mathrm{in}\,\alpha_r^{\,i}, \quad \text{with} \quad
    \alpha_r = \left(\frac{r_\mathrm{out}}{r_\mathrm{in}}\right)^{1/N_r},
\end{equation}
so that $\Delta r_i = r_{i-1/2}(\alpha_r - 1)$. We set $N_r = 128\log_{10}(r_\mathrm{out}/r_\mathrm{in})$,
i.e. $128$ cells per decade, giving $\alpha_r - 1 \simeq 1.8\%$.

The outer radius is $r_\mathrm{out} = 10^4~\unit{AU}$, large enough to enclose the initial cloud
and to keep the outer boundary dynamically irrelevant. The final inner radius is
$r_\mathrm{in} = 10^{-3}~\unit{AU}$, well inside the expected protostellar radius
\citep{vaytetProtostellarBirthAmbipolar2018,ahmadBirthMagnetizedLowmass2025}; it is not used
from the start, but reached progressively through the radial restarts described in
Sect.~\ref{subsec: restart}. In the polar direction we use $N_\theta = 128$ uniform
cells over $[0,\pi]$, giving $r\Delta\theta \simeq 0.025\,r$, comparable to the radial
resolution. In the azimuthal direction we use $N_\varphi = 64$ cells over $[0,\pi]$ rather than
$128$ cells over $[0,2\pi]$, which gives the same spatial resolution
$r\sin\theta\,\Delta\varphi \simeq 0.05\,r\sin\theta$ while suppressing odd azimuthal modes,
for the reason explained in Sect.~\ref{subsec: restart}. The grid is summarised in
Table~\ref{tab: grid}.

The $\sin\theta$ factor in the azimuthal resolution makes the cells vanishingly thin close to
the polar axis, which would cripple the time step in a region of no physical interest. We
therefore use the azimuthal grid coarsening of \textsc{Idefix}
\citep{lesurIDEFIXVersatilePerformanceportable2023}, merging cells in $\varphi$ by a factor
\begin{equation}
    f_\mathrm{coarse} = 2^n, \quad
    n = \max\left(0, \left\lfloor -\log_2\left(\sin\theta\right)\right\rfloor\right),
\end{equation}
No coarsening is applied for $\pi/6 \leq \theta \leq 5\pi/6$; closer to the axis, the factor
doubles each time $\sin\theta$ is halved (a factor of 2 for $\theta < \pi/6$, 4 for
$\theta \lesssim 14.5^\circ$, and so on, symmetrically in the southern hemisphere). Since
$2^n\sin\theta > 1/2$ by construction, the azimuthal spacing
$f_\mathrm{coarse}\, r\sin\theta\,\Delta\varphi$ never falls below
$r\Delta\varphi/2 = r\Delta\theta$ anywhere in the grid.
%-------------------------------------------------------------------
\subsection{Restart procedure}
\label{subsec: restart}

Following the collapse over seven decades in radius, and from axisymmetry to a fully
three-dimensional protostar--disc system, is made tractable by a sequence of restarts in which
the extent of the domain --- but not its relative resolution --- is changed. They are detailed in
Appendix~\ref{app: restart}, and we only summarise them here.

The inner radius starts at $1~\unit{AU}$ and is reduced by one decade at a time down to
$10^{-3}~\unit{AU}$, the new cells being filled by an extrapolation that preserves the
solenoidal constraint: the first reduction is performed as the first Larson core is about to
form, the second at the onset of the second collapse, and the third as the second Larson core
is about to form. The first collapse is computed in axisymmetry, and the simulation is restarted
in three dimensions shortly before the formation of the first Larson core, with an $m=2$
density perturbation to seed non-axisymmetric structures. Until just before the second collapse
it is moreover restricted to the northern hemisphere, which enforces equatorial symmetry and
keeps the first core at the centre of the grid. Finally, the azimuthal domain is restricted to
$\varphi \in [0,\pi]$, which suppresses odd azimuthal modes: in a full azimuthal domain, an
$m=1$ mode systematically develops and ejects the protostar from the centre of the grid. The
higher-resolution run of Table~\ref{tab: simulations} is obtained by restarting the reference run
on a refined grid.

%-------------------------------------------------------------------
\subsection{Boundary conditions}
\label{subsec: boundary}

Boundary conditions are imposed through two layers of ghost cells. In MHD only the
hydrodynamical variables and the magnetic field components tangential to the boundary need to
be specified; the normal component follows from the divergence-free constraint.

\begin{itemize}
    \item \textbf{Azimuthal direction.} Periodic. In addition, the normal component
    $B_\varphi$ must be explicitly forced to take the same value on both sides, otherwise
    round-off errors accumulate over long integrations and generate a spurious toroidal field.

    \item \textbf{Polar axis.} We use the regularised axis boundary condition of
    \citet{zhuGlobalEvolutionAccretion2018}, which assigns a single value to fields defined
    multiple times on the singular axis. Since our domain spans $\varphi \in [0,\pi]$, it
    reduces to a reflection with $\rho$, $v_r$ and $B_r$ copied and $v_\theta$, $v_\varphi$,
    $B_\theta$, $B_\varphi$ reversed.

    \item \textbf{Mid-plane.} During the half-domain phase
    (Appendix~\ref{subsubsec: polar_restart}), a reflective condition is applied at
    $\theta = \pi/2$: $v_\theta$, $B_r$ and $B_\varphi$ are reversed in the ghost cells, the
    other fields copied.

    \item \textbf{Outer radius.} Outflow: zero-gradient for all fields, except the radial
    velocity, which is set to zero if it points inwards, so that no material can be accreted
    through the outer boundary.

    \item \textbf{Inner radius.} A plain outflow condition would impose a vanishing pressure
    gradient at $r_\mathrm{in}$ and could not support a hydrostatic core. We therefore use a
    modified outflow condition, identical to the outer one except for the density, which is set
    in the ghost cells to the constant value $\rho_\mathrm{in}$ obtained by integrating the mass
    flux through the inner boundary over time. The sphere $r < r_\mathrm{in}$ then behaves as a
    homogeneous body of density $\rho_\mathrm{in}$, against which a hydrostatic core can build
    up.
\end{itemize}

The inner boundary nevertheless remains the most delicate part of the setup, since the
zero-gradient condition on $B_\theta$ and $B_\varphi$ generates artificial currents. We assess
its consequences in Sect.~\ref{subsec: limitation}.

%-------------------------------------------------------------------
\subsection{Internal limiters}
\label{subsec: limiters}

Three limiters are required to keep the integration tractable, or physically meaningful, in
the low-density polar cavities that open above the protostar. All three act in regions that we
show below to be dynamically disconnected from the disc and the protostar, but we describe them
in full since they are not innocuous for the polar regions.

\subsubsection{Diffusivity limiter}
\label{subsubsec: diffusivity_limiter}

\begin{figure}[t]
    \centering
    \includegraphics[width=\columnwidth]{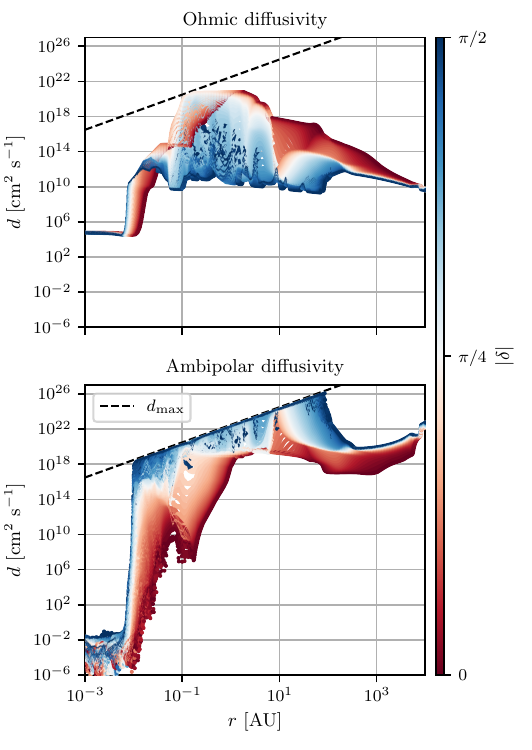}
    \caption{Ohmic (\textit{top}) and ambipolar (\textit{bottom}) diffusivities as a function of
    spherical radius, after the protostar formation. Colours show
    $|\delta| = |\pi/2 - \theta|$, the angle to the equatorial plane. The dashed line is the
    diffusivity limiter $\eta_\mathrm{max}$, computed for
    $\Delta t_\mathrm{hyp} \sim 3\times 10^{-6}~\unit{yr}$.}
    \label{fig: diffusivity_cap}
\end{figure}

Even with super-time-stepping, the ratio $\zeta$ of the hyperbolic to the parabolic time step
can exceed $10^6$ in a protostellar collapse. Following
\citet{vaytetProtostellarBirthAmbipolar2018} and
\citet{mauxionModelingSecularEvolution2024}, we cap each diffusivity at
\begin{equation}
    \eta_\mathrm{max} = C_\mathrm{diff} \frac{\Delta l^2}{2\Delta t_\mathrm{hyp}},
\end{equation}
with $\Delta l$ the local cell size and $C_\mathrm{diff} = 100$, so that a hyperbolic time step
is covered in $\sim 2\sqrt{C_\mathrm{diff}} = 20$ sub-steps instead of $\sim 200$. Where the cap
is active, diffusion still dominates the local dynamics, so the physical behaviour is
preserved. On a logarithmic spherical grid $\eta_\mathrm{max} \propto r^2$, which is most
restrictive at small radii; two effects mitigate this, namely the reduction of the hyperbolic
time step once the protostar forms, and the drop of the diffusivities caused by dust
vaporisation at high temperature. As Fig.~\ref{fig: diffusivity_cap} shows, the cap does not
affect the Ohmic diffusivity at all at this stage, and reduces the ambipolar diffusivity only in
the polar regions, outside the disc and the protostar.
% Appendix~\ref{app: diffusivity_limiter}.

\subsubsection{Alfv\'en speed and mean-free-path limiters}

\begin{figure}[t]
    \centering
    \includegraphics[width=\columnwidth]{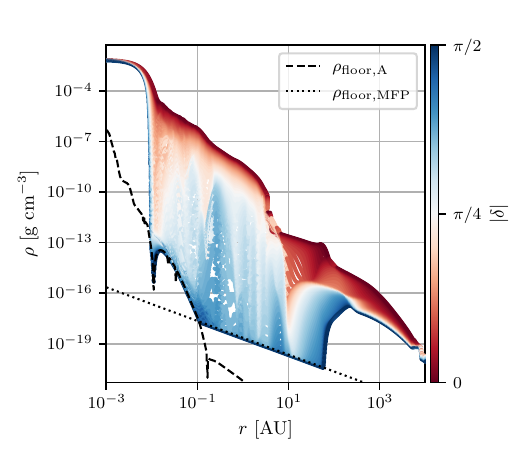}
    \caption{Density as a function of spherical radius after the protostar formation, with
    colours showing $|\delta| = |\pi/2-\theta|$, illustrating the two density floors. The dashed
    line is the floor $\rho_\mathrm{floor,A}$ associated with the Alfv\'en speed limiter, here
    for $v_\mathrm{A,ref} = 4000~\unit{AU\,yr^{-1}}$; the dotted line is the floor
    $\rho_\mathrm{floor,MFP}$ associated with the mean-free-path limiter. Both are active only in the polar cavities.}
    \label{fig: density_floors}
\end{figure}

Polar cavities open above the protostar after the formation of the first Larson core, with
densities falling below $10^{-20}~\unit{g\,cm^{-3}}$. Two problems follow. First, the
Alfv\'en speed $v_\mathrm{A} = B/\sqrt{4\pi\rho}$ reaches several
$10^3~\unit{km\,s^{-1}}$ there. Through the CFL condition, these few cells
set a very restrictive time step over the whole domain, and the cost is paid at every step of the
simulation in order to follow wave propagation in a region that holds a negligible fraction of
the mass, and that is not the object of this study. We therefore cap it at
\begin{equation}
    v_\mathrm{A,max} = v_\mathrm{A,ref}\,\frac{r}{1~\unit{AU}},
\end{equation}
which is enforced by imposing the density floor
$\rho_\mathrm{floor,A} = B^2/(4\pi v_\mathrm{A,max}^2)$. The linear radial dependence has the
same origin as for the diffusivity cap: on a logarithmic grid the CFL condition scales as
$\Delta t \propto r/v_\mathrm{A,max}$, so a larger Alfv\'en speed can be tolerated at larger
radii for the same time-step restriction. We use
$v_\mathrm{A,ref} \sim 90~\unit{AU\,yr^{-1}}$ during the first-core phase, raised to
$\sim 4000~\unit{AU\,yr^{-1}}$ once the protostar has formed and the sound speed has reduced
the hyperbolic time step by two orders of magnitude.

Second, at such densities the particle mean free path exceeds the cell size and the fluid
description breaks down. We impose a second density floor,
\begin{equation}
    \rho_\mathrm{floor,MFP} = \frac{\mu_\mathrm{mol} \mH}{\sqrt{2}\,\sigma\, l_\mathrm{max}},
    \quad \text{with} \quad l_\mathrm{max} = \frac{\Delta l}{100},
\end{equation}
where $\sigma = 4.58\times 10^{-15}~\unit{cm^2}$ is the collisional cross section estimated from
the van der Waals radius of $\mathrm{H_2}$ \citep{weastCRCHandbookChemistry1981}. This floor
scales as $r^{-1}$, whereas $\rho_\mathrm{floor,A}$ follows the magnetic field, so their
relative importance changes as the collapse proceeds and depending on the local radius: the two are comparable during the
first-core phase, while once the second core has formed, the Alfvén speed floor becomes the dominant one in the inner regions
and the mean-free-path floor becomes the dominant one beyond $\sim 0.1~\unit{AU}$. Figure~\ref{fig: density_floors} shows both at that stage. They are active only in the
polar cavities, well outside the disc and the protostar; both are applied at fixed momentum and
angular momentum, and the total mass they add over the whole simulation stays below
$10^{-6}~M_\odot$, entirely negligible compared with the mass of the protostar and the disc.

\subsubsection{Temperature floor}
\label{subsubsec: temperature_floor}

In the same cavities, the barotropic law returns temperatures below $\sim 100~\unit{K}$, whereas
the gas there would be directly irradiated by the protostar. This does not create numerical problems,
but it does feed the resistivity tables with temperatures at which dust has not been
vaporised, and therefore with spuriously high resistivities. We therefore impose, once the
protostar has formed and \emph{only} for the computation of the resistivities, a temperature
floor
\begin{equation}
    T_\mathrm{floor} = T_\mathrm{floor,ref}\left(\frac{r}{0.01~\unit{AU}}\right)^{-1/2},
\end{equation}
with $T_\mathrm{floor,ref} = 4000~\unit{K}$, close to the protostellar surface temperature. The
$r^{-1/2}$ scaling follows from radiative equilibrium with the protostar, $F \propto r^{-2}$ and
$F \propto T^4$. Since the resistivities are insensitive to temperature below
$\sim 1000~\unit{K}$, the floor is not applied where $T_\mathrm{floor}$ falls below that value,
i.e. beyond $0.16~\unit{AU}$. This prescription assumes an optically thin cavity and is
therefore optimistic; its impact on the magnetic field evolution is discussed in
Sect.~\ref{subsec: limitation}.

%-------------------------------------------------------------------
\subsection{Simulations}
\label{subsec: simulations}

\begin{table}[t]
    \centering
    \caption{Simulations discussed in this paper. 1MsunHLLD is the reference run.}
    \begin{tabular}{l|c|c}
        \hline\hline
        Name & Resolution & Solver \\
        \hline
        1MsunHLLD & Table~\ref{tab: grid} & HLLD \\
        1MsunHLLDx2 & $\times 2$ in $\theta,\varphi$ & HLLD \\
        1MsunHLL & Table~\ref{tab: grid} & HLL \\
        \hline
    \end{tabular}
    \label{tab: simulations}
\end{table}

All the simulations discussed here share the initial conditions of
Sect.~\ref{subsec: initial_conditions} and the grid of Table~\ref{tab: grid}, and differ only in
their angular resolution and in the Riemann solver, as listed in
Table~\ref{tab: simulations}. The reference run, 1MsunHLLD, is followed for
$\sim 64~\unit{yr}$ after the formation of the protostar. The convergence run 1MsunHLLDx2 is
about an order of magnitude more expensive and is followed for $\sim 17~\unit{yr}$ only. The
HLL run, followed for $\sim 70~\unit{yr}$, is presented in Appendix~\ref{app: hll}.

%-------------------------------------------------------------------
\subsection{Angular momentum budget}
\label{subsec: angular_momentum_budget}

The analysis presented in Sects.~\ref{sec: angular_momentum_equilibrium} and~\ref{sec: transport_mechanisms} relies on a decomposition of the angular
momentum budget, which we define here. Throughout, $\averagePhi{X}$ denotes the azimuthal
average of a field $X$ over the azimuthal domain $\varphi \in [0,\pi]$,
\begin{equation}
    \averagePhi{X}(r,\theta) = \frac{1}{\pi}\int_0^{\pi} X(r,\theta,\varphi)\,\mathrm{d}\varphi ,
\end{equation}
and $\delta X = X - \averagePhi{X}$ its non-axisymmetric part. We write $s = r\sin\theta$ for
the cylindrical radius, $\Omega = v_\varphi/s$ for the angular velocity, and use the subscript
$\mathrm{p}$ for the poloidal part of a vector field.

Taking the azimuthal average of the conservation equation for the axial angular momentum, and
assuming a steady state, gives
\begin{equation}
    \begin{aligned}
    \averagePhi{\rho\bm{v}_\mathrm{p}}\cdot \nabla (s^2\Omega) = - \nabla \cdot
    \Big( & s\averagePhi{\rho \delta v_{\varphi} \bm{v}_\mathrm{p}}
    - \frac{s}{4\pi}\averagePhi{B_{\varphi} \bm{B}_\mathrm{p}} \\
    & + \frac{s}{4\pi G} \averagePhi{g_{\varphi} \bm{g}_\mathrm{p}} \Big),
    \end{aligned}
    \label{eq: angular_momentum_equilibrium}
\end{equation}
where $\bm{g} = -\nabla\Phi$. The left-hand side is the advection of the fluid along the
specific angular momentum gradient, i.e. the accretion or decretion of mass; the right-hand
side gathers the three stresses that transport angular momentum, namely the Reynolds stress,
which measures the correlation of the velocity fluctuations; the Maxwell stress, which requires
a toroidal field correlated with a poloidal one; and the gravitational stress, which requires
non-axisymmetric density structures.

Because our disc is thick and our grid is spherical, we evaluate the divergence in spherical
rather than in the cylindrical coordinates more commonly used in the disc literature. The
right-hand side then splits into six local torques, one radial and one meridional for each
stress:
\begin{equation}
    \begin{aligned}
        \mathcal{H}_r &= -\frac{1}{r^2}\partial_r \left( r^3\averagePhi{\rho \delta v_{\varphi} v_r}\right), \\
        \mathcal{H}_\theta &= -\frac{1}{\sin \theta}\partial_\theta \left( \sin^2\theta\averagePhi{\rho \delta v_{\varphi} v_\theta}\right), \\
        \mathcal{M}_r &= \frac{1}{4\pi r^2}\partial_r \left( r^3\averagePhi{B_{\varphi} B_r}\right), \\
        \mathcal{M}_\theta &= \frac{1}{4\pi \sin\theta}\partial_\theta \left( \sin^2\theta\averagePhi{B_{\varphi} B_\theta}\right), \\
        \mathcal{G}_r &= -\frac{1}{4\pi G r^2}\partial_r \left( r^3\averagePhi{g_{\varphi} g_r}\right), \\
        \mathcal{G}_\theta &= -\frac{1}{4\pi G \sin\theta}\partial_\theta \left( \sin^2\theta\averagePhi{g_{\varphi} g_\theta}\right),
    \end{aligned}
    \label{eq: torques}
\end{equation}
with $\mathcal{H} = \mathcal{H}_r + \mathcal{H}_\theta$ the total Reynolds torque, and
similarly for the Maxwell torque $\mathcal{M}$ and the gravitational torque $\mathcal{G}$.
Separating the radial from the meridional component matters here: in a thick, self-gravitating
disc, angular momentum is transported vertically as well as radially, and this distinction is
lost in the vertically integrated, one-dimensional $\alpha$-disc description
\citep{shakuraBlackHolesBinary1973,pringleAccretionDiscsAstrophysics1981}.

%%%%%%%%%%%%%%%%%%%%%%%%%%%%%%%%%%%%%%%%%%%%%%%%%%%%%%%%%%%%%%%%%%%%%%%%%%%%%%%%%%%%%%%%%%%
%%%%%%%%%%%%%%%%%%%%%%%%%%%%%%%%%%%%%%%%%%%%%%%%%%%%%%%%%%%%%%%%%%%%%%%%%%%%%%%%%%%%%%%%%%%
\section{Overview of the protostar--disc system}
\label{sec: overview}

All the results presented in this and the following sections are drawn from the reference run 1MsunHLLD
(Table~\ref{tab: simulations}), which follows the protostar--disc system for its first
$\sim 64~\unit{yr}$ of evolution. We take the formation of the protostar, defined as the time
at which the peak density first exceeds $3\times 10^{-5}~\unit{g\,cm^{-3}}$, as the origin of time
throughout, and use the azimuthal average $\averagePhi{X}$ and the non-axisymmetric deviation
$\delta X$ introduced in Sect.~\ref{subsec: angular_momentum_budget}.

\begin{figure}[t]
    \centering
    \includegraphics[width=\columnwidth]{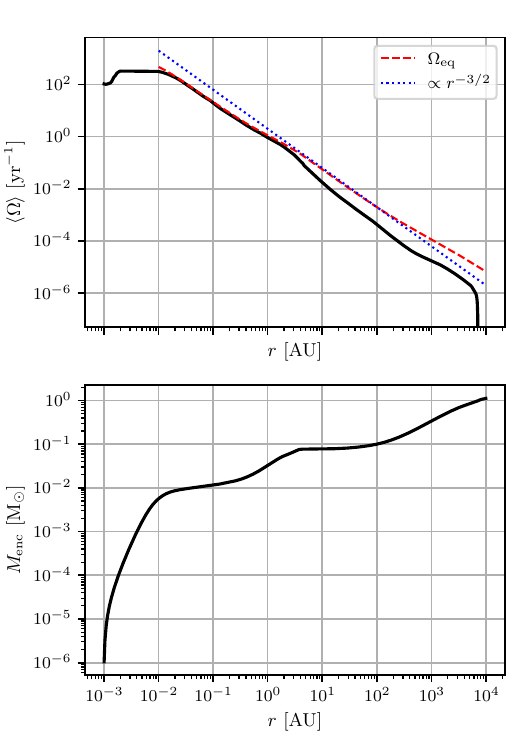}
    \caption{Azimuthally averaged angular velocity at the mid-plane (\textit{top}) and
    enclosed mass (\textit{bottom}) as a function of radius, $10~\unit{yr}$ after the protostar
    formation. The red dashed line
    is the equilibrium angular velocity $\Omega_\mathrm{eq} = \sqrt{GM_\mathrm{enc}/r^3}$
    computed from the enclosed mass, and the blue dotted line a Keplerian
    $\Omega \propto r^{-3/2}$ profile.}
    \label{fig: omega_menc}
\end{figure}

\begin{figure*}[p]
    \centering
    \includegraphics[width=\textwidth]{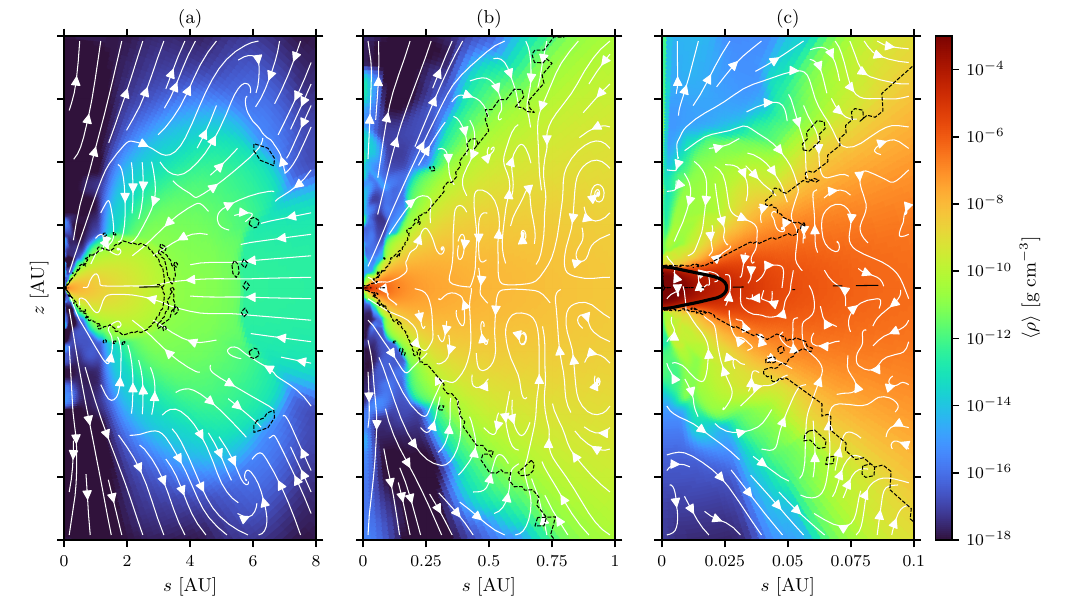}
    \caption{Density in the $s-z$ plane at three successive scales, $\sim 12~\unit{yr}$ after
    the protostar formation. In each panel, the vertical ticks have the same spacing as the
    radial ones. Colours show the density
    and white lines the poloidal velocity field, both averaged in the azimuthal direction. 
    The dashed lines indicate the boundaries of the disc. The
    black contour in panel (c) marks the protostar surface, defined as the region where
    the density exceeds $3\times 10^{-5}~\unit{g\,cm^{-3}}$.}
    \label{fig: zoom_density}
\end{figure*}

\begin{figure*}[p]
    \centering
    \includegraphics[width=\textwidth]{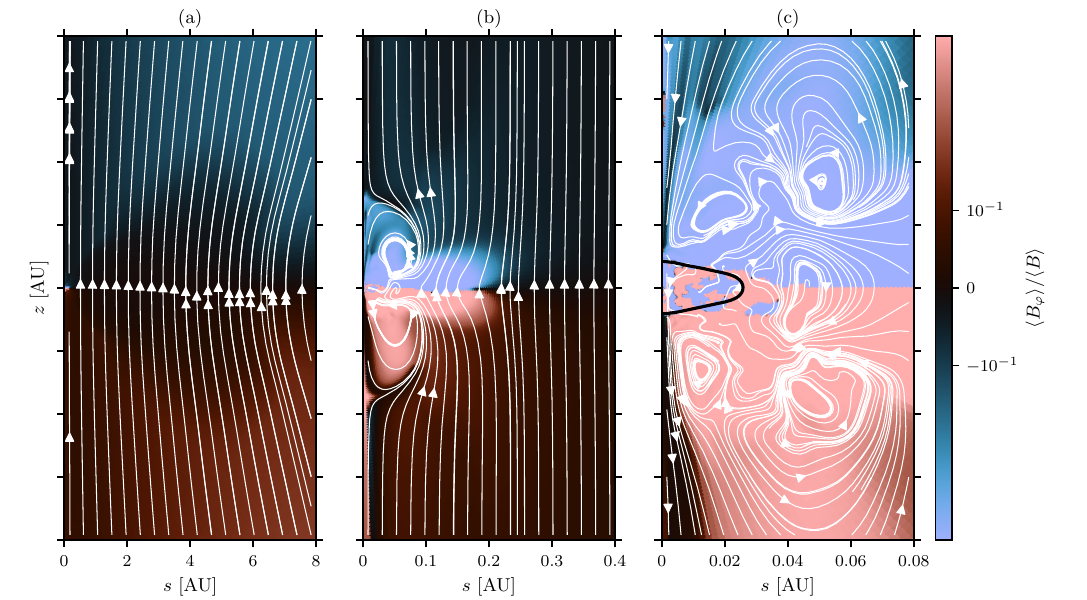}
    \caption{Magnetic field structure in the $s-z$ plane at three successive scales,
    $\sim 12~\unit{yr}$ after the protostar formation. In each panel, the vertical ticks have the
    same spacing as the radial ones. Colours show the ratio
    $\averagePhi{B_\varphi}/\averagePhi{B}$ and white lines the poloidal magnetic field, both
    averaged in the azimuthal direction. The black contour in panel (c) marks the
    protostar surface.}
    \label{fig: zoom_bphi}
\end{figure*}

Before turning to angular momentum transport, we describe the system that the collapse
produces. We distinguish three structures throughout. The \emph{protostar} is the region where
the density exceeds $3\times 10^{-5}~\unit{g\,cm^{-3}}$, the threshold also used to define the
origin of time. The \emph{disc} is the rotationally supported region that surrounds it, defined
as the region where gravity is balanced by the centrifugal and pressure forces; its surface is
drawn as a dashed or dotted contour in the maps that follow. The
\emph{envelope} is the remaining, infalling gas. We then describe the rotation of the protostar
and the disc, the flow that feeds them, the regimes of the magnetic field, the spiral structure
that develops in the disc, and the way the protostar itself evolves.
\subsection{Protostar and disc}

The collapse produces two distinct structures, which the rotation profile separates cleanly.
Figure~\ref{fig: omega_menc} shows the mid-plane angular velocity and the enclosed mass as a
function of radius. Inside $\sim 10^{-2}~\unit{AU}$, $\averagePhi{\Omega}$ is constant: the
protostar is in solid-body rotation, at $\Omega_\mathrm{protostar} \sim 200~\unit{rad\,yr^{-1}}$,
i.e. with a rotation period of $\sim 11$~days. The enclosed mass flattens
at the same radius, giving a protostellar mass of $\sim 8\times 10^{-3}~M_\odot$.

Outside this radius, and out to $\sim 2~\unit{AU}$, the angular velocity follows the
equilibrium profile $\Omega_\mathrm{eq} = \sqrt{GM_\mathrm{enc}/r^3}$: this is a
rotationally supported circumstellar disc, which formed from the first Larson core and is
therefore already in place when the protostar appears. The profile is noticeably shallower
than Keplerian in the outer part of the disc, because the mass of the disc is not negligible compared with that of the central object: the
enclosed mass reaches $\sim 8\times 10^{-2}~M_\odot$ by the
disc outer edge, so the disc is an order of magnitude more massive than the protostar
it surrounds, and it is largely self-gravitating. This has a direct consequence for what
follows --- gravitational torques cannot be dismissed a priori in such a disc. The disc outer
edge, where the rotation profile departs from $\Omega_\mathrm{eq}$ and where the enclosed mass
profile flattens, lies at $\sim 2-4~\unit{AU}$.

\subsection{Density structure and flow}

Figure~\ref{fig: zoom_density} shows the density and the poloidal velocity field at three
successive scales. At the largest scale (Fig.~\ref{fig: zoom_density}a), the disc is embedded in the infalling envelope and
fed through a pseudo-disc, a flattened structure supported vertically by thermal and magnetic
pressure but not radially \citep{galliCollapseMagnetizedMolecular1993}, along which material is funnelled towards the centre. The disc
itself is hot and geometrically thick, with an aspect ratio $H/R \sim 1$: it is supported
vertically by its thermal pressure, as also found by \citet{ahmadFormationLowmassProtostars2024}.
The thin-disc approximation therefore does not apply, and vertical and radial transport are a
priori comparable.

Two features of the flow are worth noting at this stage, since both will reappear in the
angular momentum budget. First, at the AU scale (Fig.~\ref{fig: zoom_density}b) the poloidal velocity field is organised into
large axisymmetric meridional eddies, corresponding to a systematic vertical motion of the gas
rather than to purely radial accretion. Second, at the protostellar scale (Fig.~\ref{fig: zoom_density}c) the mean flow is
disordered, with small eddies, and the non-axisymmetric fluctuations are strong
(Sect.~\ref{subsec: nonaxi}); low-density polar cavities have opened above and below the disc, from which
material is ejected. Over the duration of our simulation, the mass carried by these ejections
is negligible; they may nonetheless carry away a non-negligible amount of angular momentum, and
their mass is likely to become significant on longer timescales
\citep{machidaFirstTwoThousand2019}. Since they are launched from precisely the regions where the
density floors and the Alfv\'en speed limiter are active (Sect.~\ref{subsec: limiters}), we do
not analyse them further here.

\subsection{Magnetic field regimes}

\begin{figure}[t]
    \centering
    \includegraphics[width=\columnwidth]{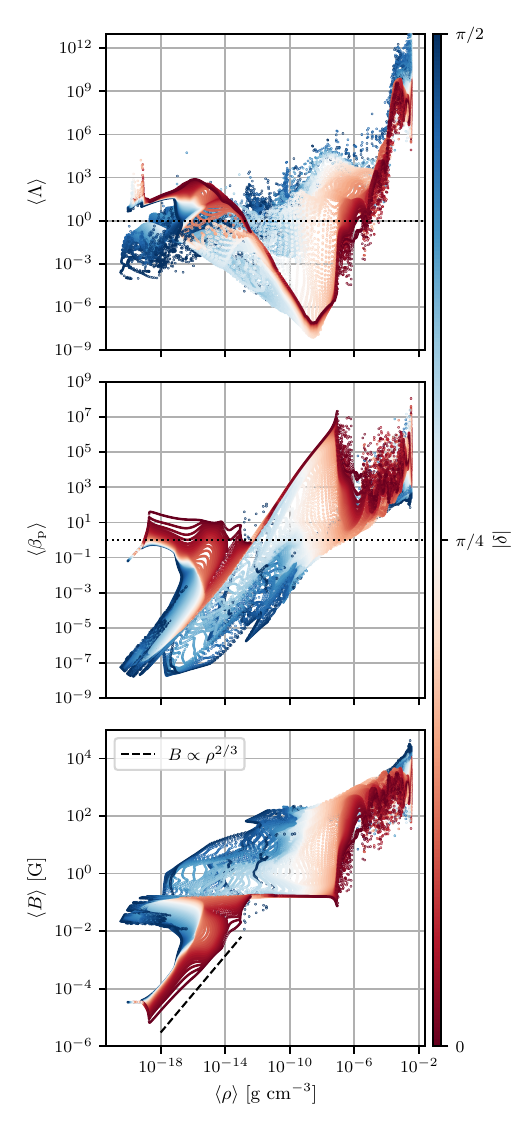}
    \caption{Azimuthally averaged Els\"asser number $\Lambda$ of Eq.~\eqref{eq: elsasser}
    (\textit{top}), plasma beta $\beta_\mathrm{p}$ (\textit{middle}) and magnetic field strength
    (\textit{bottom}) as a function of density, $\sim 12~\unit{yr}$ after the protostar
    formation. Colours show $|\delta| = |\pi/2-\theta|$, the angle to the equatorial plane, so
    that red points sample the disc and blue points the polar regions. The dotted lines mark
    $\Lambda = 1$ and $\beta_\mathrm{p} = 1$, and the dashed line is the ideal-MHD scaling
    $B \propto \rho^{2/3}$.}
    \label{fig: b_rho}
\end{figure}
\begin{figure*}[t]
    \centering
    \includegraphics[width=\textwidth]{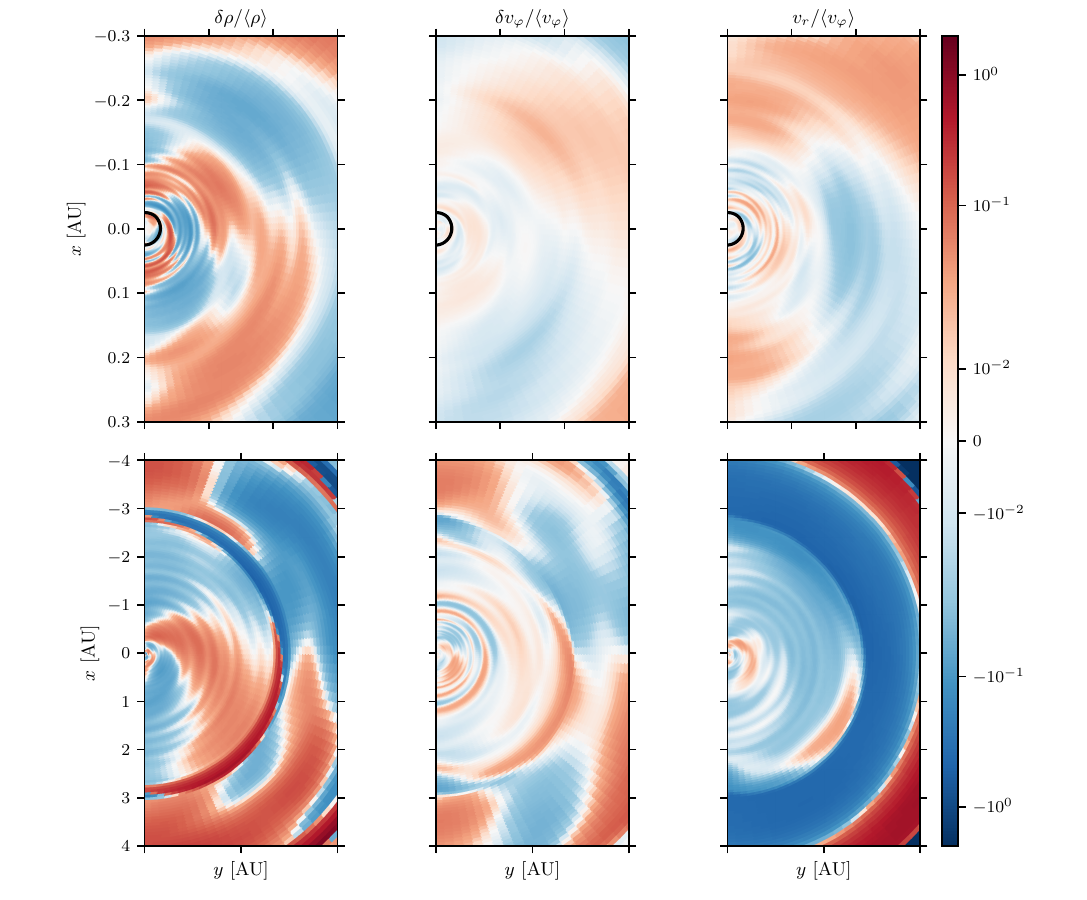}
    \caption{Non-axisymmetric structure of the disc in the $x-y$ plane at $z=0$, at two
    scales: $0.3~\unit{AU}$ (\textit{top}) and $4~\unit{AU}$ (\textit{bottom}). From left to
    right: the relative density perturbation $\delta\rho/\averagePhi{\rho}$, the relative
    azimuthal velocity perturbation $\delta v_\varphi/\averagePhi{v_\varphi}$, and the radial
    velocity normalised by the azimuthal one, $v_r/\averagePhi{v_\varphi}$. Only half of the
    disc is shown, since the azimuthal domain spans $\varphi \in [0,\pi]$
    (Sect.~\ref{subsec: restart}). The black contour marks the protostar surface.}
    \label{fig: topview}
\end{figure*}
\begin{figure*}[t]
    \centering
    \includegraphics[width=\textwidth]{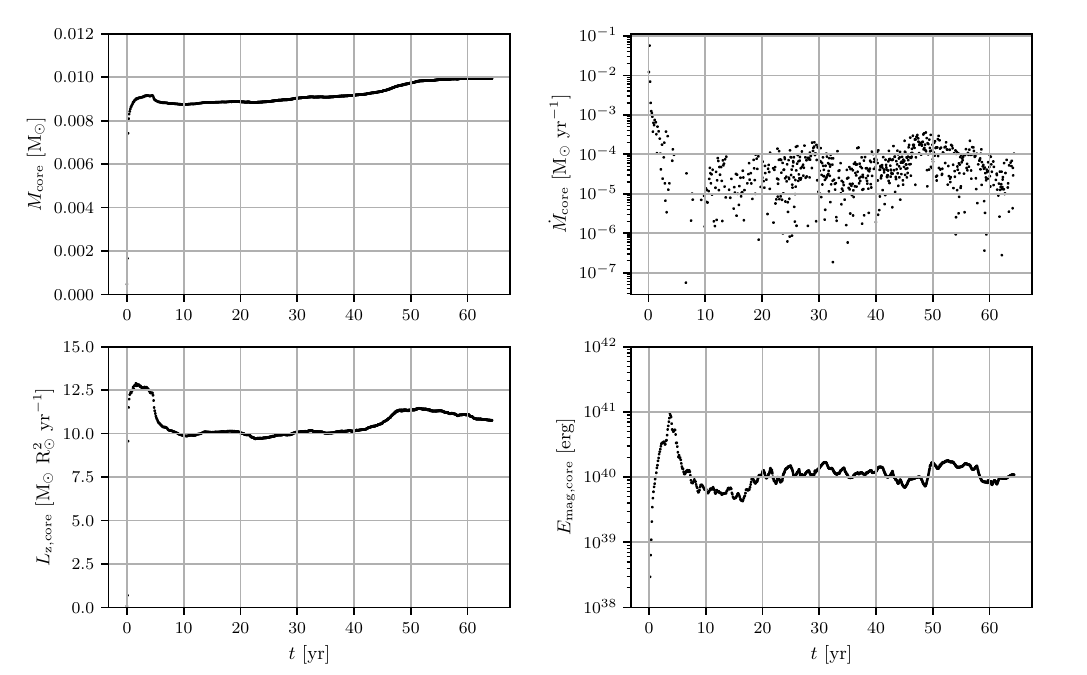}
    \caption{Evolution of the protostar properties as a function of time after its formation:
    mass (\textit{top left}), accretion rate (\textit{top right}), axial angular momentum
    (\textit{bottom left}) and magnetic energy (\textit{bottom right}). All quantities are
    integrated over the region enclosed by the protostar surface, and the accretion rate is the time
    derivative of the protostar mass.}
    \label{fig: core_evolution}
\end{figure*}
The magnetic field enters the angular momentum budget only through the Maxwell stress, which
requires a toroidal field correlated with a poloidal one. Such a stress can only develop where
the field is coupled to the gas and wound up by its rotation, and we first identify where this
is the case. Figures~\ref{fig: zoom_bphi} and~\ref{fig: b_rho} reveal two distinct regimes.

Immediately around the disc, out to a few AU, the field is nearly vertical and
almost purely poloidal (Fig.~\ref{fig: zoom_bphi}a): ambipolar diffusion has decoupled the field
from the gas, which can neither drag nor wind it up. On larger scales ($\gtrsim 7-8$~AU), in the
envelope, the field is coupled to the gas again, and the toroidal component generated by the
rotation of the envelope is no longer negligible. The $B-\rho$ relation follows the ideal-MHD scaling
$B \propto \rho^{2/3}$ at low density, but saturates at $\sim 0.1~\unit{G}$ over several decades
in density near the midplane, from $\sim 10^{-14}$ to $\sim 10^{-7}~\unit{g\,cm^{-3}}$
(Fig.~\ref{fig: b_rho}). This plateau is the signature of ambipolar diffusion decoupling the
field from the gas: the field is no longer amplified by compression, and it cannot be wound up
by the differential rotation of the disc. The coupling is measured more directly by the
Els\"asser number,
\begin{equation}
    \Lambda = \frac{v_\mathrm{A}^2}{\left(\eta_\Omega + \eta_\mathrm{A}\right)\Omega},
    \label{eq: elsasser}
\end{equation}
with $v_\mathrm{A} = B/\sqrt{4\pi\rho}$ the Alfv\'en speed, which compares the rate at which the
rotation of the gas winds up the field with the rate at which Ohmic and ambipolar diffusion
dissipate it. In the mid-plane, $\Lambda$ falls below unity at $\rho \gtrsim 10^{-13}~\unit{g\,cm^{-3}}$
and reaches $\sim 10^{-8}$ near $10^{-8}~\unit{g\,cm^{-3}}$ (Fig.~\ref{fig: b_rho}, top panel): this
region is a magnetic dead zone, in which the Maxwell stress is expected to vanish.

Below $\sim 0.1~\unit{AU}$, the picture reverses. The temperature is high enough to vaporize the dust grains, 
causing the resistivities to drop sharply \citep{marchandChemicalSolverCompute2016,borderiesDustEvolutionChemisputtering2025} and the field to recouple to the gas:
$\Lambda$ rises steeply with density and exceeds unity again above $\sim 10^{-7}~\unit{g\,cm^{-3}}$.
The differential rotation of the inner disc winds
the poloidal field into a strong toroidal component: $|B_\varphi|/B$ approaches unity and
changes sign across the mid-plane (Fig.~\ref{fig: zoom_bphi}b), while the field
strength rises steeply with density to reach $3-10~\unit{kG}$ in the protostar region. Closer to
the protostar (Fig.~\ref{fig: zoom_bphi}c), the field structure becomes disordered and appears
turbulent, a point we return to in Sect.~\ref{subsec: mri}.

The plasma beta $\beta_\mathrm{p}$ gives a complementary piece of information. While $\Lambda$ indicates whether
the field is coupled to the gas, $\beta_\mathrm{p}$ indicates how large a stress a coupled field can exert:
normalised by the gas pressure, the Maxwell stress is bounded by
$|B_r B_\varphi|/4\pi P \leq B^2/8\pi P = 1/\beta_\mathrm{p}$. In the mid-plane, $\averagePhi{\beta_\mathrm{p}}$
reaches $\sim 10^{7}$ in the dead zone and drops by several orders of magnitude in the inner disc
(Fig.~\ref{fig: b_rho}, middle panel), so that the recoupled field is also strong enough to
exert a significant stress there.
The system therefore separates into a magnetically dead outer disc and a magnetically active
inner disc, with the transition set by the recoupling radius near $0.1~\unit{AU}$. We should
consequently expect the Maxwell torque to be confined to the inner region, and some other
mechanism to be responsible for the transport further out. Section~\ref{sec: transport_mechanisms} confirms
this expectation quantitatively.

\subsection{The non-axisymmetric structure of the disc}
\label{subsec: nonaxi}

The description so far has treated the disc as axisymmetric. It is not, and the departures are
large. Figure~\ref{fig: topview} shows the mid-plane in the $x-y$ plane, at the scale of the
inner disc and at the scale of the disc as a whole.

The disc is threaded by spiral arms, visible in the density across the whole disc and, in the
sub-au region, in the azimuthal and radial velocities as well. The dominant azimuthal
wavenumber is $m=2$, as expected for a self-gravitating disc. Their signature in $v_r$ is the
one that will matter later: the arms do not merely modulate the density, they drive a radial
velocity of order a percent of the orbital velocity, alternating in sign across each arm. Both
the Reynolds and the gravitational stresses of Eq.~\eqref{eq: torques} are, by construction,
produced by departures from axisymmetry --- the first requires correlated velocity
fluctuations, the second a non-axisymmetric mass distribution --- so these arms are the natural
candidates to carry them. We take that possibility up in
Sect.~\ref{sec: transport_mechanisms}.

Two further features are worth noting. In the innermost disc, inside $\sim 0.2~\unit{AU}$, a
set of concentric rings appears in the density and the radial velocity but not in the azimuthal
velocity, which identifies them as sound waves launched by the protostar--disc interaction
rather than as a rotating pattern. In the outer disc, beyond $\sim 1~\unit{AU}$, the structure
becomes less obviously spiral and more disordered, with strong perturbations concentrated near
the outer edge; this is the most variable part of the disc.

\begin{figure*}[t]
    \centering
    \includegraphics[width=\textwidth]{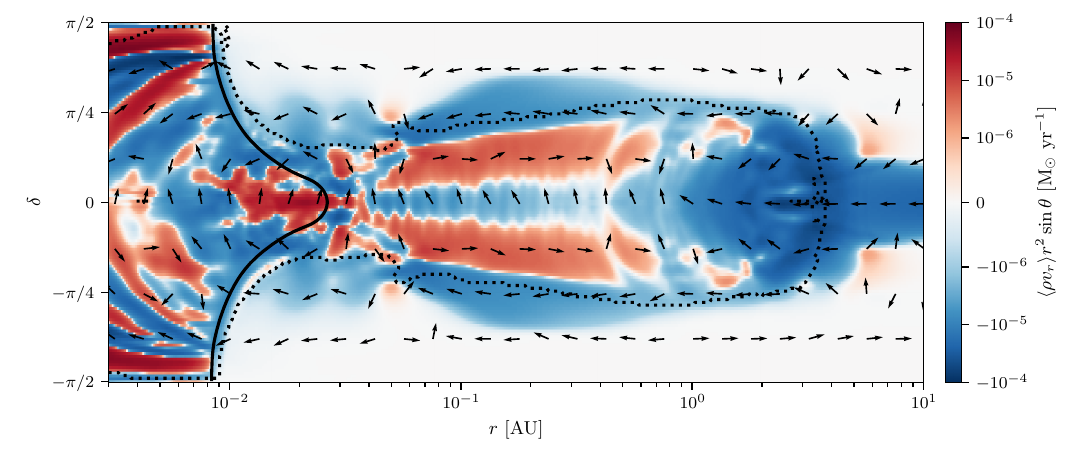}
    \caption{Azimuthally averaged radial mass flux, rescaled by $r^2\sin\theta$ to account for
    the geometry, averaged in time between $t=10~\unit{yr}$ and $t=40~\unit{yr}$. Blue
    denotes accretion and red decretion; the arrows show the direction of the poloidal mass
    flux. The vertical axis is $\delta = \pi/2 - \theta$, the angle to the equatorial plane.
    The solid black line marks the protostar surface and the dotted contour the disc surface,
    defined as the region where gravity is balanced by the centrifugal and pressure forces.}
    \label{fig: mass_flux}
\end{figure*}

\begin{figure*}[t]
    \centering
    \includegraphics[width=\textwidth]{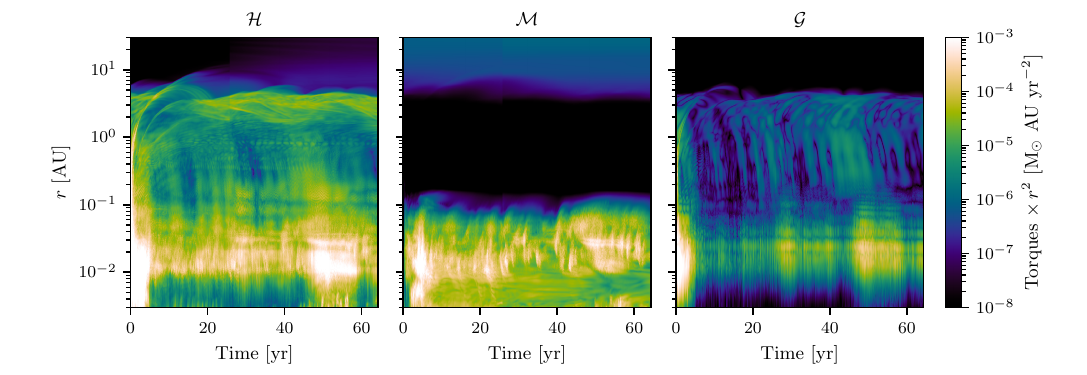}
    \caption{Amplitude of the Reynolds (\textit{left}), Maxwell (\textit{middle}) and
    gravitational (\textit{right}) torques as a function of time and radius, rescaled by
    $r^2$. Each term is first averaged in azimuth, and its absolute value is then averaged in
    the polar direction between $\theta = \pi/4$ and $\theta = 3\pi/4$, so that the maps
    measure the total transport activity at each radius.}
    \label{fig: torque_time}
\end{figure*}

\subsection{Evolution of the protostar}
\label{subsec: protostar_evolution}

Figure~\ref{fig: core_evolution} shows the time evolution of the mass, accretion rate, axial
angular momentum and magnetic energy of the protostar. The mass evolves in two stages. After a
brief adjustment in the first few years it grows only slowly, from $\sim 8.8\times 10^{-3}$ to
$\sim 9.2\times 10^{-3}~M_\odot$ over the following thirty years, with an accretion rate --- computed as the time derivative of the protostar mass --- that
drops from its initial value by three orders of magnitude and then fluctuates between $10^{-6}$
and $10^{-4}~M_\odot~\unit{yr^{-1}}$. Between $\sim 43$ and $\sim 52~\unit{yr}$ this quiescent
growth is interrupted by an accretion burst: the accretion rate rises to a sustained
$10^{-4}-3\times 10^{-4}~M_\odot~\unit{yr^{-1}}$, and the protostar gains
$\sim 8\times 10^{-4}~M_\odot$, nearly a tenth of its mass, in under a decade, before returning
to its earlier rate. Episodic accretion of this kind was also found by
\citet{machidaFirstTwoThousand2019}, where it is driven by gravitational torques in the disc. We do
not investigate the origin of this burst further here, and focus on the quasi-steady phase that
precedes it (Sect.~\ref{sec: angular_momentum_equilibrium}).

The angular momentum behaves quite differently. It peaks at
$\sim 13~M_\odot R_\odot^2~\unit{yr^{-1}}$ about two years after the protostar forms, then
decreases by roughly a fifth over the following ten years and remains near
$10~M_\odot R_\odot^2~\unit{yr^{-1}}$ until the burst. During this quiescent phase, the angular momentum of
the protostar thus stays nearly constant while its mass slowly increases, so that its
\emph{specific} angular momentum decreases: the angular momentum brought in by the accreted
material is extracted from the protostar. The burst reverses this for a time --- the angular momentum
rises to $\sim 11.4~M_\odot R_\odot^2~\unit{yr^{-1}}$, the material accreted during the burst
bringing in more angular momentum than the stresses remove --- after which the decline resumes.
The magnetic energy, for its part, rises steeply after formation, peaking at
$\sim 10^{41}~\unit{erg}$ around $4~\unit{yr}$, and then settles into oscillations about
$10^{40}~\unit{erg}$ for the rest of the run, without responding appreciably to the burst. That
it peaks during the early extraction phase already suggests that magnetic stresses play a part
in it.

The protostar loses specific angular momentum while continuing to accrete from a rotationally
supported disc. Neither is possible without an efficient mechanism to carry angular
momentum outwards, and the nature of that mechanism is what the rest of this paper
sets out to determine.

\section{Angular momentum equilibrium}
\label{sec: angular_momentum_equilibrium}

\begin{figure*}[t]
    \centering
    \includegraphics[width=\textwidth]{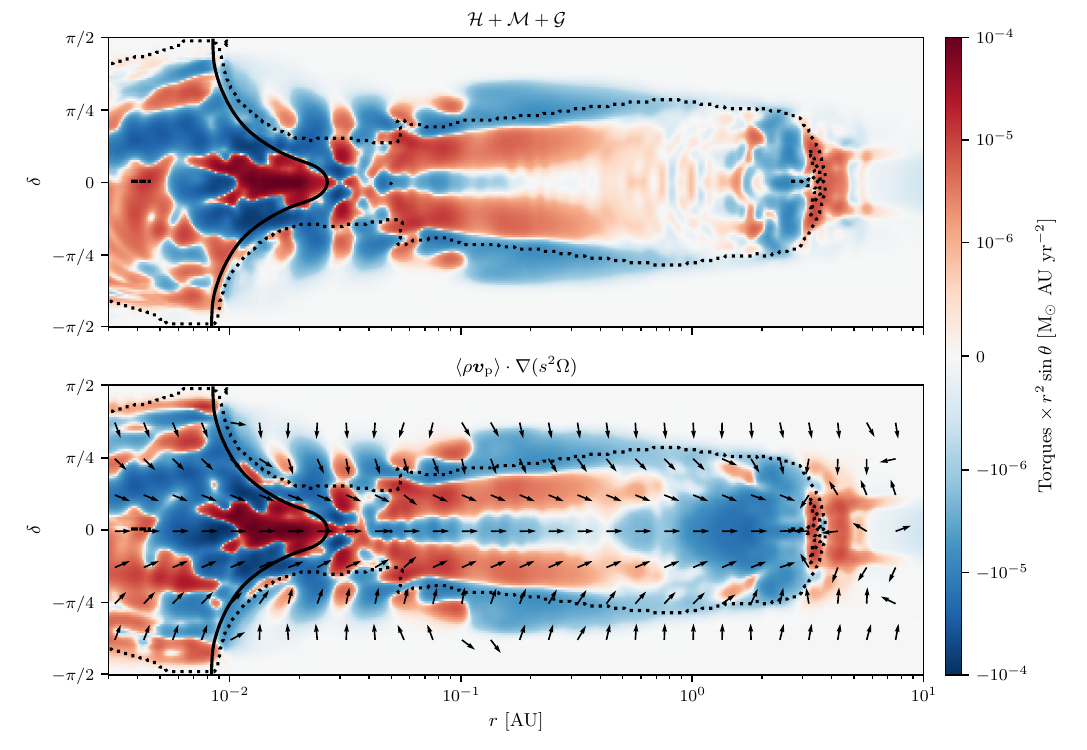}
    \caption{Total torque $\mathcal{H}+\mathcal{M}+\mathcal{G}$ (\textit{top}) and advection
    term $\averagePhi{\rho\bm{v}_\mathrm{p}}\cdot\nabla(s^2\Omega)$ (\textit{bottom}), the two
    sides of Eq.~\eqref{eq: angular_momentum_equilibrium}, averaged in time between
    $t=10~\unit{yr}$ and $t=40~\unit{yr}$ and rescaled by $r^2\sin\theta$. The arrows in the
    bottom panel show the direction of $\nabla(s^2\Omega)$, onto which the poloidal mass flux is
    projected in the advection term, i.e. the direction in which the specific angular momentum
    increases. Contours are as in
    Fig.~\ref{fig: mass_flux}.}
    \label{fig: closure}
\end{figure*}

\subsection{Where the mass actually flows}
\label{subsubsec: accretion}

Section~\ref{sec: overview} established that the protostar accretes while its specific
angular momentum decreases. The natural first question is through which parts of the disc the accreted material
travels. Figure~\ref{fig: mass_flux} shows the azimuthally and time-averaged radial mass flux
$\averagePhi{\rho v_r}$. Since the cells of the map correspond to surfaces of different sizes
depending on their radius and latitude, we multiply the flux by $r^2\sin\theta$: the quantity
shown, $r^2\sin\theta\averagePhi{\rho v_r}$, is the mass accretion rate per unit polar and
azimuthal angle, so that the same colour at different radii and latitudes corresponds to the same
mass accretion rate.

The result is not a disc that accretes as a whole. Beyond the disc outer edge, at
$r \gtrsim 3~\unit{AU}$, the envelope falls in and feeds the system through the pseudo-disc,
producing the broad inflow that dominates the map at those radii. Inside the disc, however,
between $\sim 0.05$ and $\sim 1~\unit{AU}$, the flux breaks up into vertically stacked layers
of alternating sign: accretion and decretion coexist at the same radius, at different heights
above the mid-plane. Closer in, below $\sim 0.05~\unit{AU}$, the pattern becomes irregular,
and the mid-plane just outside the protostar surface shows net decretion: material is
leaving the protostar there, even though the protostar as a whole is gaining mass
(Fig.~\ref{fig: core_evolution}).

Accretion and decretion therefore coexist at the same radius, in different layers of the
disc. This is not a detail of the flow: the two directions carry comparable fluxes, so a
description in terms of a single radial accretion rate --- the vertically integrated,
one-dimensional $\alpha$-disc picture \citep{shakuraBlackHolesBinary1973,%
pringleAccretionDiscsAstrophysics1981} --- would report a net accretion rate for a region
in which material is simultaneously moving both ways. Since a fluid element in a
rotationally supported disc can only move radially by exchanging angular momentum, this
layered pattern must reflect an equally layered angular momentum transport.

\subsection{Choice of the averaging window}

Equation~\eqref{eq: angular_momentum_equilibrium} assumes a steady state, and the maps above
are time averages, so we must first establish over which interval such a state holds.
Figure~\ref{fig: torque_time} shows the amplitude of the three torques as a function of time and
radius over the whole run.

All three share a strong transient during the first $\sim 5~\unit{yr}$, while the protostar and
its immediate environment settle after the second collapse; it coincides with the peak of the magnetic energy of the
protostar and with a strong decrease of its angular momentum near $t \sim 5~\unit{yr}$
(Fig.~\ref{fig: core_evolution}).
After $\sim 10~\unit{yr}$ the torques fluctuate about well-defined radial profiles without
systematic drift, and they continue to do so until $\sim 40~\unit{yr}$. Around $43~\unit{yr}$
the Reynolds and gravitational torques in the inner disc, between $\sim 0.01$ and
$\sim 0.03~\unit{AU}$, increase sharply, and the Maxwell torque strengthens in the same region:
this is the accretion burst described in Sect.~\ref{subsec: protostar_evolution}.

We restrict the analysis to $t = 10-40~\unit{yr}$, and all time averages below are taken over
that window. 
This window is the interval over which the mass and angular momentum of the protostar evolve slowly
and its accretion rate shows no secular trend (Fig.~\ref{fig: core_evolution}), which is
precisely the condition under which the time derivative neglected in
Eq.~\eqref{eq: angular_momentum_equilibrium} is small. During the burst that term is not
negligible --- the angular momentum of the protostar rises by some fifteen per cent in under a
decade --- and averaging across it would combine two different regimes. A weaker
episode of gravitational activity is also visible in the inner disc around $25-30~\unit{yr}$,
with a corresponding rise in the accretion rate. We have checked that neither this episode nor
the choice of window conditions our results: repeating the analysis
separately over $t = 10-25$ and $25-40~\unit{yr}$ gives qualitatively the same results --- the
same three transport regions and the same ranking of the stresses within each, as presented in
Sect.~\ref{sec: transport_mechanisms} --- so that the quasi-steady phase is well
characterised by a single average. 
% Over the burst, $t = 40-60~\unit{yr}$, the picture changes:
% the gravitational stress becomes much more dominant contribution in the inner disc, where the Maxwell
% stress dominates during the quasi-steady phase. The burst is therefore not a large fluctuation
% of the quiescent transport but a distinct transport regime, which is the stronger reason for
% treating it separately.

Figure~\ref{fig: torque_time} also already shows the single most robust feature of the
transport, which does not depend on any of the details discussed later: the Maxwell
torque is active below $\sim 0.1~\unit{AU}$, essentially absent between $\sim 0.1$ and
$\sim 7~\unit{AU}$, and non-zero again at larger radii. The gap is the dead zone identified in
Sect.~\ref{sec: overview}, seen now in the transport itself rather than in the field
structure; beyond it, in the envelope, the field is still ideal and the Maxwell torque
reappears, as discussed in Sect.~\ref{sec: transport_mechanisms}. The Reynolds torque, by contrast, is
active at all radii, with maxima near the protostar surface and near the disc outer edge, and
the gravitational torque is weaker but present throughout the disc.

\subsection{Accretion is controlled by angular momentum transport}
\label{subsec: closure}

We first verify that the two sides of Eq.~\eqref{eq: angular_momentum_equilibrium} agree.
Figure~\ref{fig: closure} shows the total torque $\mathcal{H}+\mathcal{M}+\mathcal{G}$ and the
advection term $\averagePhi{\rho\bm{v}_\mathrm{p}}\cdot\nabla(s^2\Omega)$ side by side, on the
same scale.

The two maps agree closely in sign and in structure over most of the domain, and in particular
they reproduce the same layered pattern between $0.05$ and $1~\unit{AU}$ and the same
mid-plane feature just outside the protostar surface. The agreement is best in the inner disc,
between $\sim 0.01$ and $\sim 0.3~\unit{AU}$, where the orbital time is shorter than a year, so
that the averaging window covers many orbits and fluctuations average out efficiently. Further
out the agreement degrades, for the reason one would expect: near $1~\unit{AU}$ the orbital
time approaches the length of the averaging window itself, only a few orbits are sampled, and
the steady-state assumption becomes correspondingly weaker. The region around $1~\unit{AU}$ is
also one where the total torque happens to be small, so that residual dynamical evolution is
comparatively more visible. Inside the protostar ($r \lesssim 0.01~\unit{AU}$) the two maps are
not expected to match at all, since that region is pressure- rather than centrifugally
supported and Eq.~\eqref{eq: angular_momentum_equilibrium} does not describe it.

Within those limits, the budget closes. This is a non-trivial check: the six torques of
Eq.~\eqref{eq: torques} are computed independently of the mass flux, from second-order
correlations of the velocity, magnetic and gravitational fields, so their sum reproducing the
advection map validates both the steady-state approximation in the inner disc and our
evaluation of the individual terms.

The physical content of Fig.~\ref{fig: closure} lies in the correspondence between the two
maps: where the stresses remove angular momentum, the gas accretes; where they deposit it,
the gas moves outwards. The correspondence holds layer by layer, so that the vertical
structure of the mass flux is set by the vertical structure of the transport, as anticipated
in Sect.~\ref{subsubsec: accretion}.

The mid-plane region near $0.01~\unit{AU}$ illustrates this most clearly. There the total
torque is strongly positive: the stresses deposit angular momentum, the centrifugal support
of the gas increases, and material is expelled from the protostar. The accreting layers, at the mid-plane further
out and at the upper and lower disc surfaces, are correspondingly those in which angular
momentum is extracted.

\section{Transport mechanisms}
\label{sec: transport_mechanisms}

\begin{figure*}[t]
    \centering
    \includegraphics[width=0.9\textwidth]{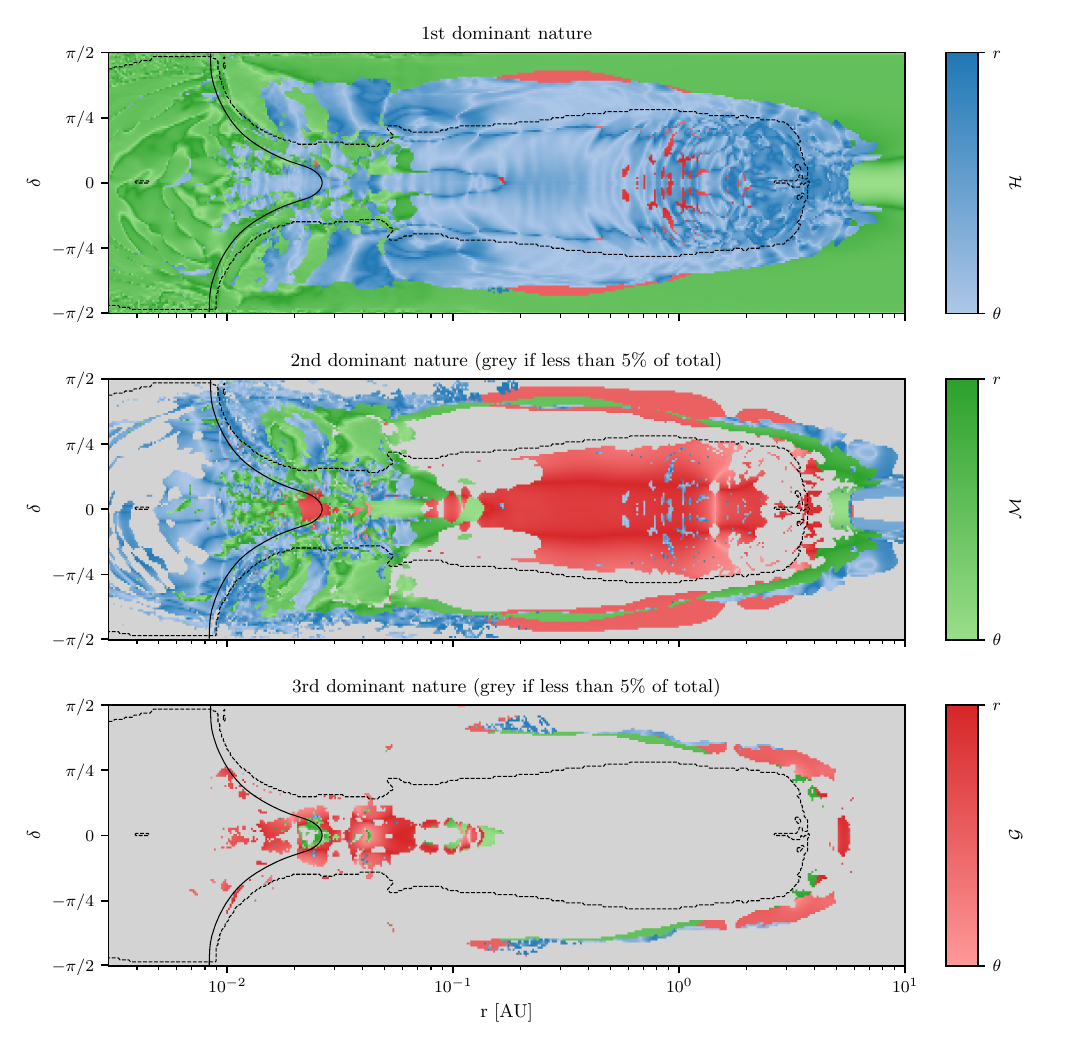}
    \caption{Relative importance of the six torques of Eq.~\eqref{eq: torques}, averaged in time
    between $t=10~\unit{yr}$ and $t=40~\unit{yr}$. At each location, the six torques are ranked
    by absolute value: the top panel shows the largest one, the middle panel the second largest,
    and the bottom panel the third largest. In the middle and bottom panels, locations where that
    torque amounts to less than 5\% of the sum of the absolute values of the six torques are
    shown in grey. Colour encodes the nature of the torque --- blue for Reynolds
    $\mathcal{H}$, green for Maxwell $\mathcal{M}$, red for gravitational $\mathcal{G}$ --- and
    shade its direction, light for the meridional component and dark for the radial one.
    Contours are as in Fig.~\ref{fig: mass_flux}.}
    \label{fig: closure_nature}
\end{figure*}

Section~\ref{sec: angular_momentum_equilibrium} established that the accretion pattern is
set by the angular momentum budget. We now decompose that budget into the six torques of
Eq.~\eqref{eq: torques} and ask which of them acts where.

\subsection{Which stress dominates where}

Figure~\ref{fig: closure_nature} ranks the six torques by amplitude at each location. Rather
than plotting six maps, we show the first, second and third most important contributions,
using colour for the nature of the stress and shade for its direction; this makes both the
identity of the dominant mechanism and the radial-versus-meridional character of the transport
readable at once. Four regions emerge.

\paragraph{The envelope and the polar regions.} Outside the disc --- beyond $\sim 3~\unit{AU}$
in the pseudo-disc, and above the disc surface --- the Maxwell torque dominates. The envelope is
still in the ideal-MHD regime, where the field remains anchored to the infalling gas, and the
toroidal component generated at these scales (Sect.~\ref{sec: overview}) carries the
associated torque. This is classical magnetic braking of the envelope, which is what delivers
low-angular-momentum material to the disc in the first place.

\paragraph{The outer disc, $1-3~\unit{AU}$.} Within the disc the Maxwell torque disappears ---
this is the dead zone --- and the Reynolds torque takes over, with radial and meridional
components of comparable importance. This is the most massive and most weakly bound part of the
disc, and also the part where the steady-state description is least secure
(Sect.~\ref{sec: angular_momentum_equilibrium}). The gravitational torque is the second
contributor, almost purely radial.

\paragraph{The middle disc, $0.1-1~\unit{AU}$.} The Reynolds torque still dominates, but now
overwhelmingly through its \emph{meridional} component. Angular momentum in this region is not
passed outwards from ring to ring, as a one-dimensional viscous model would have it, but carried
vertically, and it accumulates at intermediate latitudes, near $\delta \simeq \pm\pi/8$. The
centrifugal support rises there and material is expelled, while the mid-plane and the disc
surfaces, which lose angular momentum, both accrete. This is the three-layered pattern of
Fig.~\ref{fig: mass_flux}: accretion at the mid-plane and at the surfaces, decretion in between.
The gravitational torque is again the second contributor and again almost purely radial, so that
the two act along different directions.

\paragraph{The inner disc, $0.01-0.1~\unit{AU}$.} Inside the recoupling radius the Maxwell torque
dominates: the field has recoupled to the gas and is wound up into the
strong toroidal component observed in Fig.~\ref{fig: zoom_bphi}b. The Reynolds torque is not far behind as
the second contributor, and both are mixed in direction, with comparable radial and meridional
components. The transport here is therefore genuinely two-dimensional: angular momentum is
carried outwards in radius and upwards towards the disc surface at the same time, which is what
allows accretion to proceed along some layers while others decrete.

The gravitational torque is nowhere dominant, but it is a persistent second or third
contributor across the whole disc, and it is never negligible, as anticipated in
Sect.~\ref{sec: overview} from the disc mass.

\subsection{What drives the Reynolds torque}

\begin{figure*}[t]
    \centering
    \includegraphics[width=\textwidth]{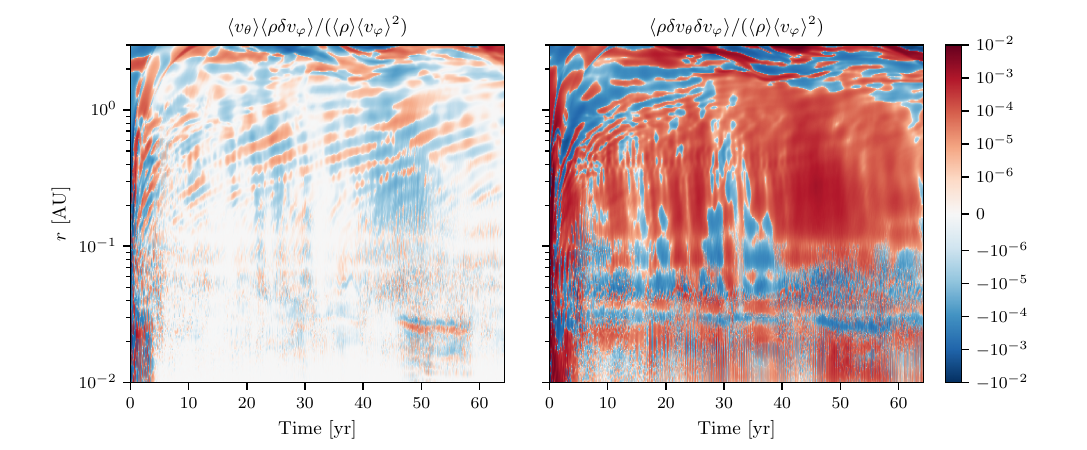}
    \caption{Decomposition of the meridional Reynolds flux, Eq.~\eqref{eq: reynolds_decomposition},
    at $\theta = 0.4\pi$, as a function of time and radius. \textit{Left:} the contribution of
    the axisymmetric flow, $\averagePhi{v_\theta}\averagePhi{\rho\,\delta v_\varphi}$.
    \textit{Right:} the contribution of the non-axisymmetric fluctuations,
    $\averagePhi{\rho\,\delta v_\theta\, \delta v_\varphi}$. Both are normalised by
    $\averagePhi{\rho}\averagePhi{v_\varphi}^2$ and shown on the same scale.}
    \label{fig: reynolds_dec}
\end{figure*}

\begin{figure}[t]
    \centering
    \includegraphics[width=\columnwidth]{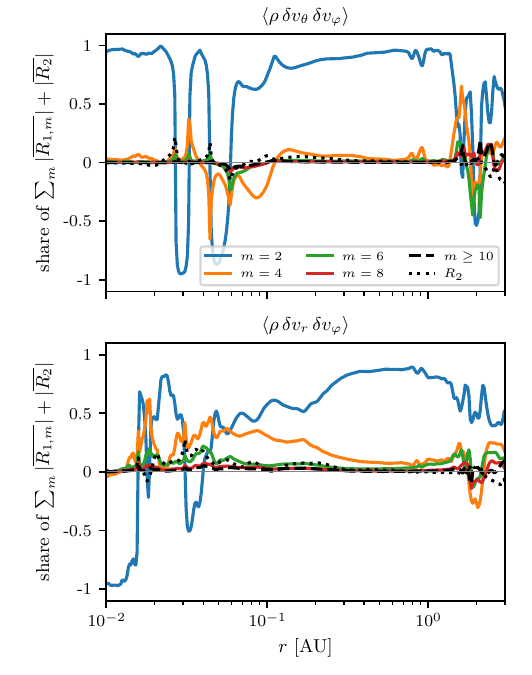}
    \caption{Azimuthal mode decomposition of the fluctuating Reynolds flux,
    Eq.~\eqref{eq: reynolds_modes}, for its meridional component
    $\averagePhi{\rho\,\delta v_\theta\,\delta v_\varphi}$ (\textit{top}) and its radial component
    $\averagePhi{\rho\,\delta v_r\,\delta v_\varphi}$ (\textit{bottom}), at $\theta = 0.4\pi$ and
    averaged in time between $t=10~\unit{yr}$ and $t=40~\unit{yr}$. Each curve shows the signed
    share of one contribution --- the azimuthal mode $m$ of $R_{1,m}$, or the density term
    $R_2$ --- that is, the contribution divided by $\sum_m |\overline{R_{1,m}}| + |\overline{R_2}|$,
    where the overline denotes the time average.}
    \label{fig: reynolds_modes}
\end{figure}

The Reynolds torque dominates the transport over most of the disc, but the label covers two
physically distinct processes. To separate them, we split the Reynolds flux into the
contribution of the mean flow and that of the non-axisymmetric fluctuations,
\begin{equation}
    \averagePhi{\rho \bm{v}\,\delta v_\varphi} = \averagePhi{\bm{v}}\,\averagePhi{\rho \delta v_\varphi}
    + \averagePhi{\rho\, \delta \bm{v}\, \delta v_\varphi} .
    \label{eq: reynolds_decomposition}
\end{equation}
The first term is the transport of the angular momentum perturbation by the axisymmetric
flow --- in practice, by the mean meridional motion --- while the second is the transport by
the non-axisymmetric fluctuations themselves. Figure~\ref{fig: reynolds_dec} shows the two terms of the meridional flux, the
component that dominates the middle disc.

The answer is unambiguous: the non-axisymmetric term exceeds the mean-flow term by roughly two
orders of magnitude throughout the region where the Reynolds torque dominates. The angular
momentum is carried by the non-axisymmetric fluctuations, not by the mean meridional
circulation. Both terms show the same pattern of outward-propagating features, but with very
different amplitudes.

The non-axisymmetric fluctuations may, however, be either the large-scale spiral arms of
Fig.~\ref{fig: topview} or small-scale turbulence. To separate the two, we decompose the
fluctuating term, for its meridional ($i=\theta$) and radial ($i=r$) components, as
\begin{equation}
    \averagePhi{\rho\,\delta v_i\,\delta v_\varphi}
    = \averagePhi{\rho}\averagePhi{\delta v_i\,\delta v_\varphi}
    + \averagePhi{\delta\rho\,\delta v_i\,\delta v_\varphi}
    = \sum_m R_{1,m} + R_2 ,
    \label{eq: reynolds_modes}
\end{equation}
where $R_{1,m}$ is the contribution of the azimuthal Fourier mode $m$ to
$\averagePhi{\rho}\averagePhi{\delta v_i\,\delta v_\varphi}$, obtained from the Fourier
coefficients of $\delta v_i$ and $\delta v_\varphi$ (Parseval's theorem), and $R_2$ the
contribution of the density fluctuations. Turbulence would spread the flux over many modes,
whereas a large-scale spiral pattern concentrates it in the lowest ones.
Figure~\ref{fig: reynolds_modes} shows that, for both components, the $m=2$ mode carries most
of the flux, typically between half and all of it, at almost all radii between $0.01$ and
$\sim 2~\unit{AU}$; the modes $m \geq 10$ and the density term $R_2$ are negligible. Beyond
$\sim 0.1~\unit{AU}$, $m=2$ also remains the dominant mode at almost all times for both
components, within and outside the averaging window, whereas inside $0.1~\unit{AU}$ it
dominates the meridional flux most of the time, but the radial flux is often dominated by the
$m=4$ mode (not shown). In the middle disc, where the Reynolds torque dominates the transport, the
fluctuations are therefore carried by the $m=2$ spiral arms rather than by turbulence. The
picture differs at both ends of the disc. In the inner disc, the higher modes carry a larger
share of the flux, mostly in the radial component, where the $m=4$ mode contributes
$\sim 20-30\%$ of the time-averaged flux between $\sim 0.015$ and $\sim 0.3~\unit{AU}$: smaller-scale
fluctuations also contribute to the transport there (Sect.~\ref{subsec: mri}). Beyond
$\sim 2~\unit{AU}$, the higher modes become dominant in places, in agreement with the more
disordered, turbulent structure of the outer disc seen in Fig.~\ref{fig: topview}. We note that,
since our azimuthal domain spans $\varphi \in [0,\pi]$, only even modes are present, and $m=2$
is the largest-scale pattern the domain allows (Sect.~\ref{subsec: restart}).

The mean-flow term is nonetheless not zero, and it is structured rather than noisy: it
organises into coherent stripes drifting outwards, on a timescale much longer than the orbital
one. This is the signature of the systematic vertical motion already visible in the poloidal
velocity field of Fig.~\ref{fig: zoom_density}. It contributes little to the angular momentum
budget, but it is a genuine feature of the flow, and we discuss its origin in
Sect.~\ref{subsec: vertical_motion}.

\subsection{Transport efficiency and the $\alpha$ parametrisation}
\label{subsubsec: alpha}

\begin{figure*}[t]
    \centering
    \includegraphics[width=\textwidth]{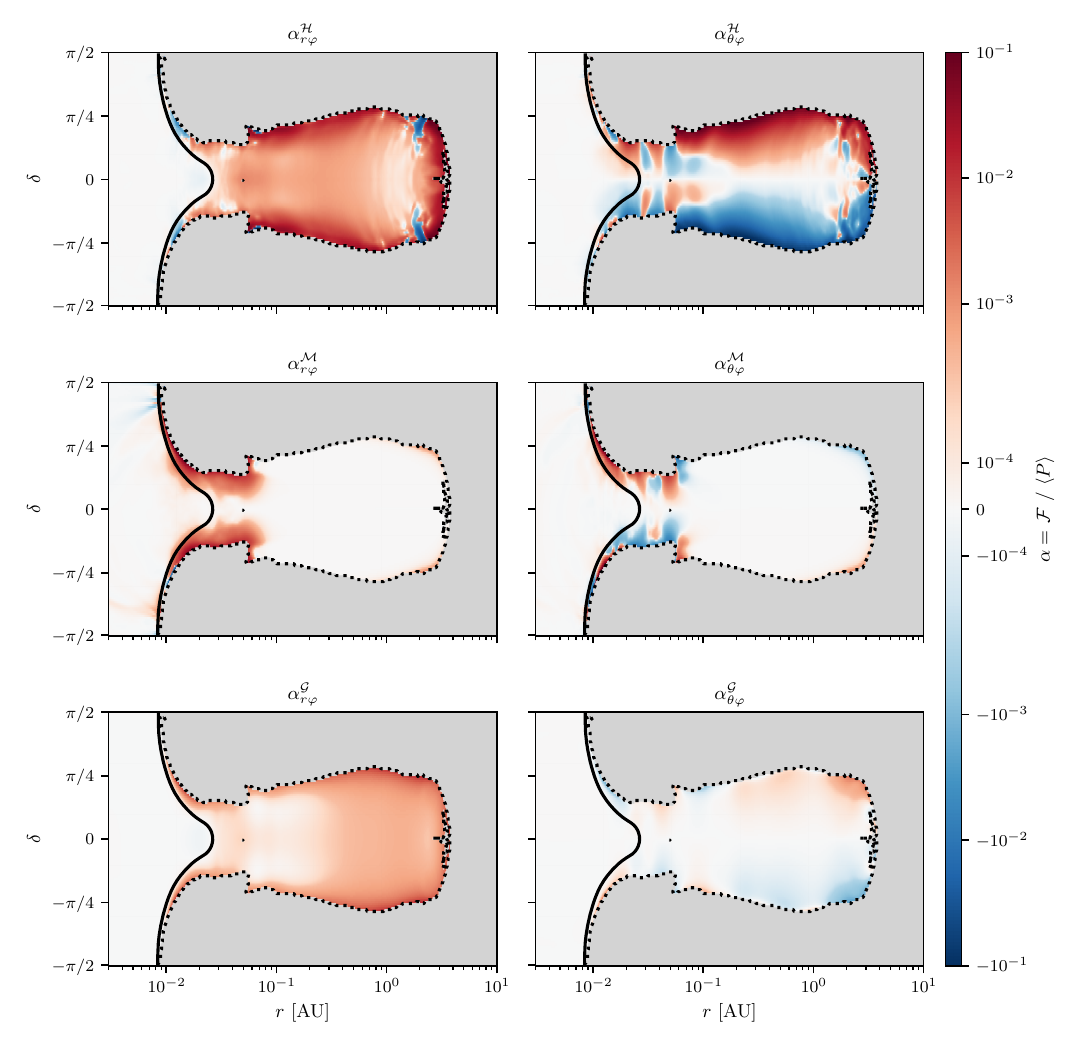}
    \caption{Dimensionless transport efficiencies $\alpha^{X}_{i\varphi}$ defined by
    Eq.~\eqref{eq: alpha_def}, averaged in time between $t=10~\unit{yr}$ and
    $t=40~\unit{yr}$. \textit{Left column:} the radial component $i=r$, the one to which the
    standard $\alpha$ prescription corresponds. \textit{Right column:} the meridional component
    $i=\theta$, which has no counterpart in that prescription. \textit{From top to bottom:} the
    Reynolds, Maxwell and gravitational stresses. The fluxes are normalised by the local gas
    pressure, and the maps are masked outside the disc surface. Contours are as in
    Fig.~\ref{fig: mass_flux}.}
    \label{fig: alpha_map}
\end{figure*}

The torques of Eq.~\eqref{eq: torques} have so far been quoted in code units, which makes them
difficult to compare with the disc literature, where transport efficiency is almost invariably
expressed as a Shakura--Sunyaev $\alpha$ \citep{shakuraBlackHolesBinary1973}.

Writing the right-hand side of Eq.~\eqref{eq: angular_momentum_equilibrium} as the divergence
of a stress, its components are
\begin{equation}
    \begin{gathered}
    \mathcal{F}^{\mathcal{H}}_{i\varphi} = \averagePhi{\rho\,\delta v_\varphi v_i}, \quad
    \mathcal{F}^{\mathcal{M}}_{i\varphi} = -\frac{1}{4\pi}\averagePhi{B_\varphi B_i}, \\
    \mathcal{F}^{\mathcal{G}}_{i\varphi} = \frac{1}{4\pi G}\averagePhi{g_\varphi g_i},
    \end{gathered}
\end{equation}
with $i = r$ or $\theta$, and we define the corresponding dimensionless efficiencies as
\begin{equation}
    \alpha^{X}_{i\varphi} = \frac{\mathcal{F}^{X}_{i\varphi}}{\averagePhi{P}},
    \qquad X \in \{\mathcal{H}, \mathcal{M}, \mathcal{G}\}.
    \label{eq: alpha_def}
\end{equation}
Two choices in Eq.~\eqref{eq: alpha_def} deserve comment. First, $\averagePhi{P}$ is the
\emph{local} gas pressure, evaluated at the same position as the flux, rather than a mid-plane
value. In a thin disc every point can be associated unambiguously with the mid-plane point
directly beneath it; in a disc as thick as ours this association is no longer straightforward,
and there is no natural choice of which mid-plane pressure to attach to a given point of the
disc. The local pressure avoids that choice. Since it falls to the floor values of
Sect.~\ref{subsec: limiters} in the polar cavities, where the ratio would diverge, we mask the
maps outside the disc surface. Second, we normalise by the gas rather than
the total pressure; the two differ appreciably only inside $\sim 0.1~\unit{AU}$, where
$\beta_\mathrm{p}$ approaches unity (Fig.~\ref{fig: b_rho}), and the choice changes $\alpha$
there by a factor of order unity.

Since Fig.~\ref{fig: alpha_map} is meant to represent the time-averaged torque maps of
Figs.~\ref{fig: closure} and~\ref{fig: closure_nature} in dimensionless form, $\alpha$ is
computed as the ratio of the time-averaged flux to the time-averaged pressure, rather than as
the time average of their ratio.

For the radial Reynolds and Maxwell components, Eq.~\eqref{eq: alpha_def} reduces to the
quantity usually denoted $\alpha$, up to the distinction between a local and a vertically
integrated normalisation. The meridional components have no counterpart in the standard
prescription, which is the point of showing them alongside.

Figure~\ref{fig: alpha_map} shows the six efficiencies over the disc. Two features stand out.

First, their magnitude. Inside $\sim 0.1~\unit{AU}$ the Maxwell stress reaches
$\alpha^{\mathcal{M}} \sim 10^{-2}-10^{-1}$, concentrated in two lobes above and below the
mid-plane at the protostar--disc interface, and falls to zero beyond that radius: the dead zone
is as sharply delimited in efficiency as it was in the torque maps. Across the dead zone the
Reynolds stress gives $\alpha^{\mathcal{H}} \sim 10^{-3}-10^{-2}$, and the gravitational stress
roughly an order of magnitude less, $\alpha^{\mathcal{G}} \sim 10^{-4}-10^{-3}$. 
% These are vigorous values by the standards of evolved protoplanetary discs, in which turbulent transport
% is generally inferred to be weaker \citep[see][]{lesurMagnetohydrodynamicsProtoplanetaryDiscs2021a};
% they describe, however, a young and massive self-gravitating disc still being fed by its
% envelope, and they are consistent with the accretion rates reported in
% Sect.~\ref{sec: overview}.

Second, the meridional efficiencies cannot be neglected compared with the radial ones:
$\alpha^{\mathcal{H}}_{\theta\varphi}$ reaches the same $\sim 10^{-2}$ as
$\alpha^{\mathcal{H}}_{r\varphi}$ over much of the dead zone. This confirms, in dimensionless
form, what the relative importance of the torques already showed
(Fig.~\ref{fig: closure_nature}): the vertical transport of angular momentum between the disc
surfaces and the mid-plane is as significant as the radial transport, and cannot be
neglected.

% Third, no single value summarises the disc. The efficiency spans some three orders of
% magnitude between the gravitational stress in the outer disc and the Maxwell stress at the
% protostar--disc interface, and it is carried by a different mechanism in each of the three
% regions. It also varies strongly with height within a single region:
% $\alpha^{\mathcal{H}}_{r\varphi}$ is weakest near the mid-plane between $0.5$ and
% $2~\unit{AU}$, and strongest towards the disc surfaces near $|\delta| \sim \pi/4$. This last
% point deserves emphasis, because the mid-plane is where a vertically integrated model
% effectively calibrates itself: an $\alpha$ measured there would understate the radial transport
% in the bulk of the disc by about an order of magnitude, before one even asks what became of the
% meridional component.

\begin{figure}[t]
    \centering
    \includegraphics[width=\columnwidth]{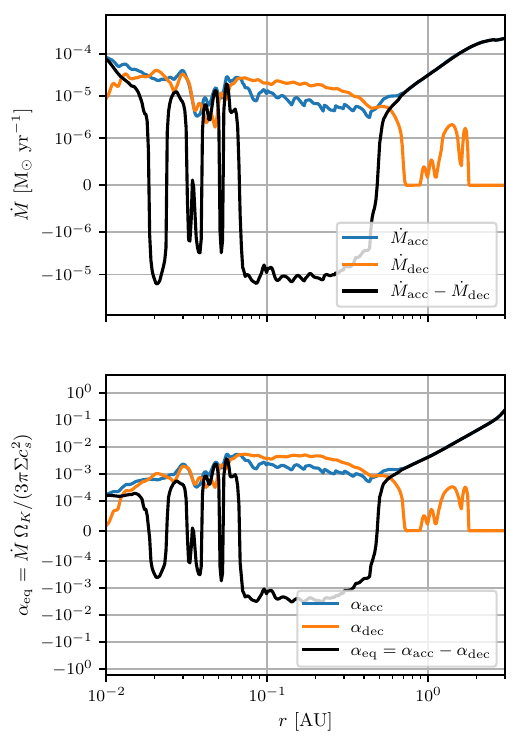}
    \caption{\textit{Top:} accreting and decreting contributions to the mass flux through the
    spherical surface of radius $r$, and their difference. At each radius
    $\dot{M}_\mathrm{acc}$ sums the cells in which the flux is inward and
    $\dot{M}_\mathrm{dec}$ those in which it is outward, both counted positively, so that their
    difference is the net inward flux. Only cells lying inside the disc are included --- the
    dotted contour of Fig.~\ref{fig: mass_flux} --- so that the polar regions are excluded.
    \textit{Bottom:} the equivalent viscosity parameter of Eq.~\eqref{eq: alpha_eq} obtained
    from each. Both panels are averaged in time between $t=10~\unit{yr}$ and
    $t=40~\unit{yr}$. Note the symmetric logarithmic scale: the net flux, and with it
    $\alpha_\mathrm{eq}$, is negative between $\sim 0.09$ and $\sim 0.35~\unit{AU}$.}
    \label{fig: alpha_acc}
\end{figure}

The maps above describe the transport locally. It is worth asking, finally, what a
one-dimensional model would infer if it were confronted only with the accretion rate that our
disc actually produces. Inverting the steady-state relation $\dot{M} = 3\pi\nu\Sigma$ with
$\nu = \alpha c_\mathrm{s}^2/\Omega_\mathrm{eq}$ gives the equivalent viscosity parameter
\begin{equation}
    \alpha_\mathrm{eq} = \frac{\dot{M}\,\Omega_\mathrm{eq}}{3\pi\Sigma c_\mathrm{s}^2},
    \label{eq: alpha_eq}
\end{equation}
that is, the number a viscous model would need in order to reproduce the measured mass flux at
each radius. Here $\Sigma(r)$ is the density integrated over $\theta$ at constant $r$ within
the disc, $c_\mathrm{s}$ the mid-plane sound speed, and $\Omega_\mathrm{eq}$ the equilibrium
angular velocity of Fig.~\ref{fig: omega_menc}. We build the viscosity from
$\Omega_\mathrm{eq}$ rather than from the Keplerian frequency because the disc carries an
appreciable fraction of the mass of the system and rotates measurably below the Keplerian rate
(Sect.~\ref{sec: overview}); using $\Omega_\mathrm{K}$ would misstate $\alpha_\mathrm{eq}$
in the outer disc, where the departure is largest. Because accretion and decretion coexist at
the same radius (Sect.~\ref{sec: angular_momentum_equilibrium}), we evaluate
Eq.~\eqref{eq: alpha_eq} separately for the inward and outward parts of the flux and for their
difference.

Figure~\ref{fig: alpha_acc} shows the result, and the two panels say different things. Taken
individually, the one-way flows are unremarkable: each corresponds to
$\alpha_\mathrm{acc}, \alpha_\mathrm{dec} \sim 10^{-3}-10^{-2}$ across the dead zone, in reasonable
agreement with the local Reynolds efficiencies of Fig.~\ref{fig: alpha_map}. The net flux,
however, is the small residual of two much larger opposing flows, and between $\sim 0.09$ and
$\sim 0.35~\unit{AU}$ that residual is negative, with $\alpha_\mathrm{eq}$ reaching
$\sim -10^{-3}-10^{-2}$. Over a third of the disc, a viscous model fitted to our mass flux would
require a negative viscosity.

The negative $\alpha_\mathrm{eq}$ has a simple explanation. A positive viscosity can only carry
angular momentum outwards, and so cannot make it accumulate at a given radius. The vertical
transport does: in the dead zone, $\alpha^{\mathcal{H}}_{\theta\varphi}$ is positive above the
mid-plane and negative below it (Fig.~\ref{fig: alpha_map}), which, with $\bm{e}_\theta$
pointing from north to south, means that angular momentum is carried from both disc surfaces
towards the disc interior, where it accumulates. A one-dimensional model, having no vertical
direction, can only mimic this accumulation with a negative viscosity.

% We stress that the objection is to the sign of $\alpha_\mathrm{eq}$, not to the sign of
% $\dot{M}$. A viscous disc can perfectly well decrete over part of its extent --- this is
% ordinary viscous spreading --- but it does so with a single change of sign, at a radius set by
% the angular momentum content of the disc, and with $\alpha > 0$ throughout. What
% Fig.~\ref{fig: alpha_acc} shows instead is a net outflow in the \emph{inner} disc, while the
% protostar accretes, and an equivalent $\alpha$ that changes sign more than once.

% Two caveats attach to all of these numbers. The normalisation by pressure is a rescaling, not a
% closure: $\alpha$ acquires its usual predictive meaning only when the transport is local and
% diffusive, which is not the case here. The Reynolds stress in the dead zone is carried by
% spiral arms whose radial extent is a sizeable fraction of the disc radius, and transport by
% self-gravitating structures is known to depart from a local viscous description when the disc
% mass is a significant fraction of the central mass
% \citep{cossinsCharacterizingGravitationalInstability2009,kratterGravitationalInstabilitiesCircumstellar2016}
% --- which, at a disc-to-protostar mass ratio of about ten, is emphatically our case. The values
% in Fig.~\ref{fig: alpha_map} should therefore be read as a measurement of how efficient the
% transport is, expressed in familiar units, and not as a prescription that could be inserted
% into a one-dimensional model.

What this leaves open is the origin of the stresses themselves, which we discuss in the next section.
% The Maxwell torque in the inner
% disc requires a mechanism to sustain the field against dissipation and to organise it; the
% spiral arms require the disc to be gravitationally unstable; and the mean vertical motion
% requires a source of vertical shear. We take up these three questions in turn in
% Sect.~\ref{sec: discussion}.

%%%%%%%%%%%%%%%%%%%%%%%%%%%%%%%%%%%%%%%%%%%%%%%%%%%%%%%%%%%%%%%%%%%%%%%%%%%%%%%%%%%%%%%%%%%%
%%%%%%%%%%%%%%%%%%%%%%%%%%%%%%%%%%%%%%%%%%%%%%%%%%%%%%%%%%%%%%%%%%%%%%%%%%%%%%%%%%%%%%%%%%%%
\section{Origin of the stresses}
\label{sec: origin}

Section~\ref{sec: transport_mechanisms} identified where each stress acts, but not what sustains
it. We now consider the origin of the two dominant ones: the magnetic stress in the inner disc
(Sect.~\ref{subsec: mri}), and the spiral arms that drive the Reynolds stress in the dead zone
(Sect.~\ref{subsec: spirals}). We then turn to the mean vertical motion responsible for the
coherent, mean-flow part of the Reynolds stress (Sect.~\ref{subsec: vertical_motion}).

%-------------------------------------------------------------------
\subsection{The magneto-rotational instability in the inner disc}
\label{subsec: mri}

\begin{figure}[t]
    \centering
    \includegraphics[width=\columnwidth]{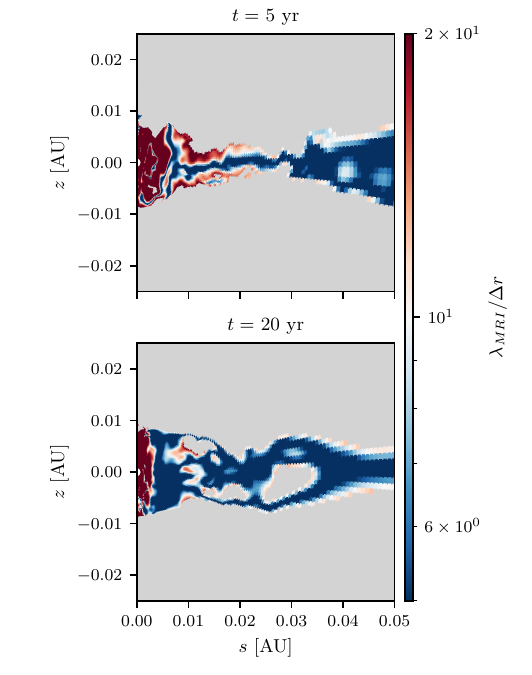}
    \caption{Resolution of the most unstable MRI mode in the inner disc, at
    $t=5~\unit{yr}$ (\textit{left}) and $t=20~\unit{yr}$ (\textit{right}). Colours show the
    number of cells per wavelength $\lambda_\mathrm{MRI}/\Delta r$, white corresponding to ten
    cells per wavelength. Regions where $\lambda_\mathrm{MRI}$ exceeds the local disc height,
    and where the instability therefore cannot develop, are shown in grey.}
    \label{fig: mri_resolution}
\end{figure}

\begin{figure}[t]
    \centering
    \includegraphics[width=\columnwidth]{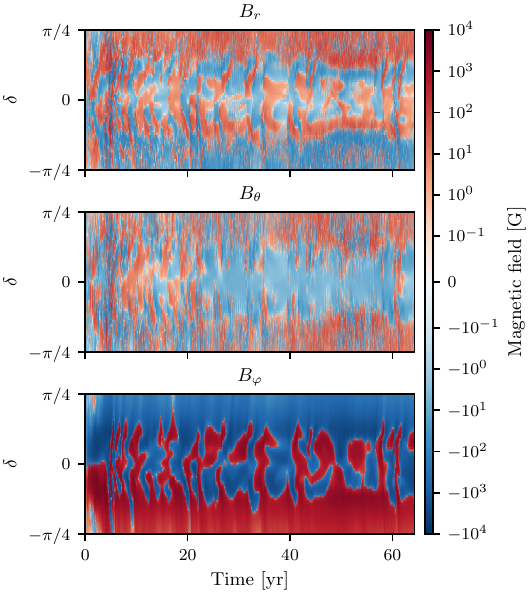}
    \caption{Butterfly diagram of the three magnetic field components at $r=0.015~\unit{AU}$,
    averaged in azimuth, as a function of time and of the angle $\delta$ to the equatorial
    plane.}
    \label{fig: butterfly}
\end{figure}

\begin{figure}[t]
    \centering
    \includegraphics[width=\columnwidth]{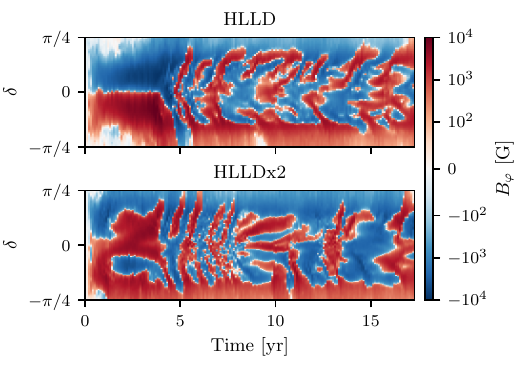}
    \caption{Butterfly diagram of the azimuthal magnetic field at $r=0.015~\unit{AU}$ in the
    1MsunHLLD (\textit{top}) and 1MsunHLLDx2 (\textit{bottom}) simulations, over the interval
    covered by both runs.}
    \label{fig: butterfly_x2}
\end{figure}

The Maxwell torque dominates the transport inside $\sim 0.1~\unit{AU}$, and
Sect.~\ref{sec: transport_mechanisms} showed that it acts radially and meridionally in
comparable measure. Simple winding of a poloidal field by differential rotation would not
produce this: it generates an ordered toroidal field and a predominantly radial stress. The
mixed, disordered character of the stress, together with the turbulent field structure visible
in Fig.~\ref{fig: zoom_bphi}c, points instead to the magneto-rotational
instability \citep[MRI,][]{balbusPowerfulLocalShear1991,hawleyLocalThreedimensionalMagnetohydrodynamic1995}.

The conditions for the MRI are met in this region. The rotation profile is differential
(Fig.~\ref{fig: omega_menc}), and the field is well coupled to the gas: this is precisely the
region inside the recoupling radius, where the Els\"asser number of Eq.~\eqref{eq: elsasser}
exceeds unity (Fig.~\ref{fig: b_rho}). The MRI also requires the field to be weak enough,
$\beta_\mathrm{p} \gtrsim 1$. Its most unstable wavelength is
\begin{equation}
    \lambda_\mathrm{MRI} = \frac{2\pi v_\mathrm{A}}{\Omega},
\end{equation}
and since the disc scale height is $H \sim c_\mathrm{s}/\Omega$, we have
$\lambda_\mathrm{MRI}/H \sim 2\pi v_\mathrm{A}/c_\mathrm{s} \propto \beta_\mathrm{p}^{-1/2}$:
requiring $\beta_\mathrm{p} \gtrsim 1$ is equivalent to requiring that $\lambda_\mathrm{MRI}$ fits
within the disc thickness. Finally, for the instability to be captured by our grid,
$\lambda_\mathrm{MRI}$ must be resolved by a sufficient number of cells.

Figure~\ref{fig: mri_resolution} shows that both conditions are satisfied over an appreciable
part of the inner disc at $t=5~\unit{yr}$, with more than ten cells per wavelength in the
innermost region. The direct evidence, however, is Fig.~\ref{fig: butterfly}: the azimuthal
field at $r=0.015~\unit{AU}$ reverses sign quasi-periodically and the reversals migrate away
from the mid-plane, producing the butterfly pattern that is the classical signature of
MRI-driven dynamo activity in a stratified disc. The radial and meridional components show the
same pattern. We conclude that the MRI operates in the inner disc and is the most likely origin
of the Maxwell torque there.

This has two consequences for the transport. First, MRI turbulence saturates the field rather
than letting it grow without bound, which is why the field strength saturates, reaching $3-10~\unit{kG}$ in the protostar region,
instead of continuing to be amplified by the shear. This saturation is not set by equipartition:
$\beta_\mathrm{p}$ remains well above unity there (Fig.~\ref{fig: b_rho}), so the field is far
from equipartition with the thermal pressure. Second, the same
turbulence feeds velocity fluctuations, and therefore contributes to the Reynolds stress. The
$m=2$ spiral pattern still carries most of the Reynolds flux inside $0.1~\unit{AU}$, but the
higher azimuthal modes contribute more there than in the middle disc, in particular to the
radial flux, which is often dominated by the $m=4$ mode (Fig.~\ref{fig: reynolds_modes}), consistent with a turbulent contribution from the
MRI. 
% The two
% are not independent contributions but two faces of the same instability.

Two caveats must be attached to this conclusion. The first is that the MRI is not numerically
converged. Figure~\ref{fig: mri_resolution} shows that by $t=20~\unit{yr}$ the region where the
instability can be contained has grown, but the number of cells per wavelength has fallen below
ten over most of it, and correspondingly the butterfly pattern weakens with time.
Figure~\ref{fig: butterfly_x2} makes the point directly: doubling the angular resolution makes
the pattern appear earlier and raises its frequency, exactly as expected when the instability is
better resolved. The MRI in our simulations is therefore under-resolved, and the amplitude of
the Maxwell torque in the inner disc should be regarded as a lower limit rather than a
converged measurement.

The second is that the result depends on the Riemann solver. In the more diffusive HLL run, no
butterfly pattern develops at the same resolution, and the field is instead amplified into an
ordered toroidal configuration by the shear alone (Appendix~\ref{app: hll},
Fig.~\ref{fig: hll_mri}) --- a difference already documented for MRI simulations by
\citet{flockHighorderGodunovSchemes2010}. This is the
main reason we adopt HLLD here. It also means that the \emph{quantitative} partition between
Maxwell and Reynolds stresses in the inner disc is resolution- and solver-dependent. In both
runs, however, the Maxwell torque is the dominant contribution inside $0.1~\unit{AU}$: this
follows from the recoupling of the field and its winding by the differential rotation, which
are set by the non-ideal MHD and not by the Riemann solver.

%-------------------------------------------------------------------
\subsection{The spiral arms}
\label{subsec: spirals}

\begin{figure}[t]
    \centering
    \includegraphics[width=\columnwidth]{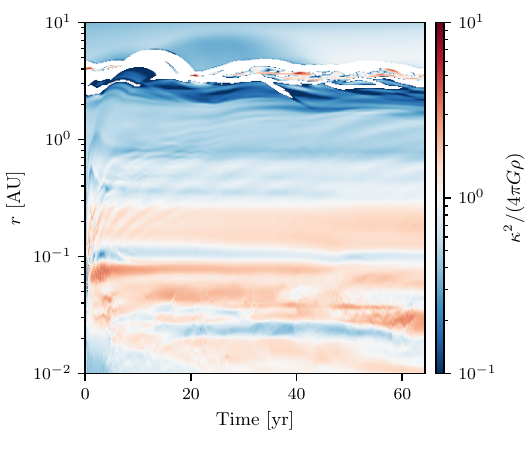}
    \caption{Three-dimensional Toomre parameter $Q_\mathrm{3D} = \kappa^2/(4\pi G\rho)$ at the
    mid-plane, as a function of time and radius. White corresponds to $Q_\mathrm{3D}=1$. The
    white band near the disc outer edge marks the region where the epicyclic frequency squared
    is negative and $Q_\mathrm{3D}$ is undefined.}
    \label{fig: toomre}
\end{figure}

Outside the recoupling radius, the field is decoupled and the transport is carried by the
Reynolds stress, which Sect.~\ref{sec: transport_mechanisms} traced to the spiral arms. The
question is then why the disc is spiral in the first place, and the natural candidate is
gravitational instability --- as anticipated in Sect.~\ref{sec: overview}, where the disc was
found to be an order of magnitude more massive than the protostar.

The standard diagnostic is the Toomre parameter
$Q = c_\mathrm{s}\kappa/(\pi G \Sigma)$ \citep{toomreGravitationalStabilityDisk1964}, with
$\kappa$ the epicyclic frequency and $\Sigma$ the surface density. Our disc is thick, so a
vertically integrated criterion is not appropriate, and we use instead the three-dimensional
form
\begin{equation}
    Q_\mathrm{3D} = \frac{\kappa^2}{4\pi G \rho},
\end{equation}
which for a thin Keplerian disc reduces to $Q_\mathrm{3D} = \sqrt{\pi/8}\,Q$ and therefore
yields a comparable stability threshold near unity.

Figure~\ref{fig: toomre} shows $Q_\mathrm{3D}$ at the mid-plane as a function of time and
radius, and the three regions of Sect.~\ref{sec: transport_mechanisms} reappear in it.
Inside $\sim 0.03~\unit{AU}$, the disc starts marginally unstable and then stabilises, with
$Q_\mathrm{3D}$ rising well above unity after $\sim 7~\unit{yr}$ as the protostar grows and
comes to dominate the local epicyclic frequency. This matches the early, transient
gravitational activity seen in the first years of Fig.~\ref{fig: torque_time}, and explains why
the gravitational torque retreats to a secondary role in the inner disc once the system has
settled.

Between $\sim 0.1$ and $\sim 1~\unit{AU}$ --- the dead zone, where the Reynolds torque takes
over --- $Q_\mathrm{3D}$ hovers close to unity throughout the simulation. The disc there is
marginally stable: unstable enough to sustain the spiral structure of
Fig.~\ref{fig: topview}, but not unstable enough to fragment, and indeed no fragmentation
occurs in any of our runs. In discs with a full thermal evolution, such a marginal state is
usually maintained by a balance between the heating of the gas in the spiral shocks and its
radiative cooling
\citep{laughlinSpiralModeSaturation1997,cossinsCharacterizingGravitationalInstability2009,kratterGravitationalInstabilitiesCircumstellar2016}.
That mechanism is absent here. With a barotropic equation of state the gas does heat as it is
compressed, its temperature rising monotonically with density, but only through that
compression: there is no dissipative heating in the spiral shocks, and no radiative cooling.
The pressure support that this compressional heating provides as the disc becomes denser is the
likely reason $Q_\mathrm{3D}$ settles close to unity.
% That the spirals are simultaneously the product of marginal instability and the agent of the
% dominant torque is the central point of this subsection: in the dead zone, the disc regulates
% its own accretion gravitationally, in the absence of any magnetic help.

Beyond $\sim 2~\unit{AU}$, $Q_\mathrm{3D}$ drops to its lowest values. This is the outer edge,
where the disc meets the infalling envelope, and where Fig.~\ref{fig: topview} shows
perturbations that are strong but no longer organised into clean spirals. It is also where the
Reynolds torque acquires a radial component comparable to its meridional one
(Fig.~\ref{fig: closure_nature}). The picture is consistent: the most gravitationally unstable
part of the disc is also the most dynamically active, and the least well described by a
steady state.

\begin{figure}[t]
    \centering
    \includegraphics[width=\columnwidth]{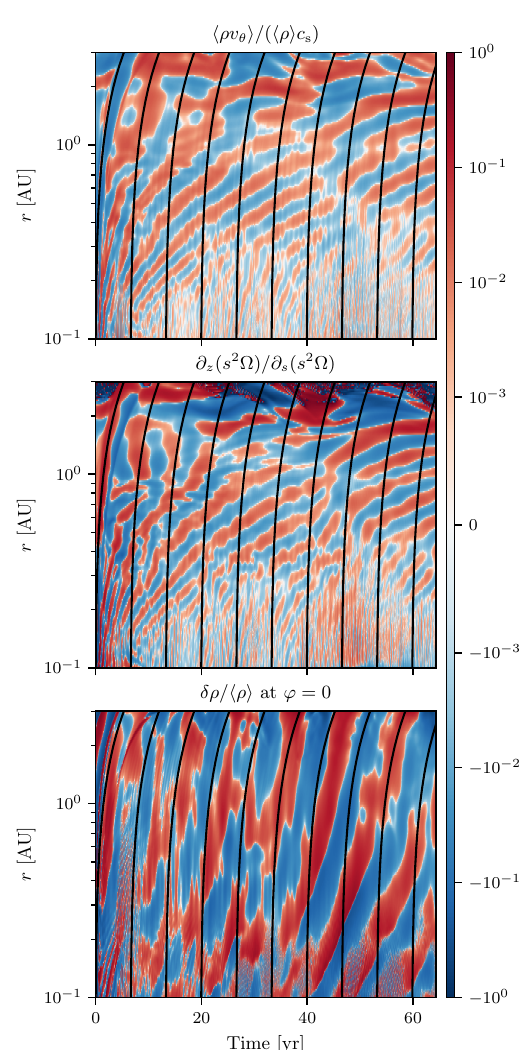}
    \caption{Characterisation of the mean vertical motion, as a function of time and radius.
    \textit{Top:} azimuthally averaged, mass-weighted meridional velocity normalised by the
    sound speed, averaged between $\theta = 0.4\pi$ and $\theta = 0.5\pi$.
    \textit{Middle:} the local ratio $\epsilon = \partial_z(s^2\Omega)/\partial_s(s^2\Omega)$
    of vertical to radial shear, over the same angular range. \textit{Bottom:} the azimuthal
    density perturbation at $\varphi = 0$ and $\theta = 0.4\pi$, tracing the spiral pattern.
    Black lines are sound-speed characteristics, spaced by the orbital time at $1~\unit{AU}$.}
    \label{fig: vsi}
\end{figure}

\subsection{Vertical motion in the disc}
\label{subsec: vertical_motion}

The decomposition of Sect.~\ref{sec: transport_mechanisms} left a residue: the mean-flow
contribution to the Reynolds stress is two orders of magnitude smaller than the fluctuating
one, but it is coherent rather than noisy, and it corresponds to the large meridional eddies
already visible in Fig.~\ref{fig: zoom_density}. Although it is not important for the angular
momentum budget, a systematic vertical circulation matters for other reasons --- dust settling
and mixing in particular --- and it is worth identifying.

Figure~\ref{fig: vsi} characterises this motion. The mean meridional velocity reaches a few
percent of the sound speed and organises into structures that propagate outwards, slowly: their
tracks in the $(t,r)$ plane are much shallower than the sound-speed characteristics drawn in
black. The middle panel shows the ratio $\epsilon$ of vertical to radial shear over the same
region. Its spatial and temporal structure closely matches that of the vertical velocity, and
the two are anti-correlated: the gas moves against the vertical shear. The amplitude of
$\epsilon$ is a few to ten percent.

This correlation seems to relate this motion to the vertical shear instability
\citep{nelsonLinearNonlinearEvolution2013,lesurHydroMagnetohydroDustGas2023}, and the slow
outward propagation is consistent with the recent analysis of
\citet{ogilvieVerticalShearInstability2025}. The motion is odd with respect to the mid-plane,
i.e. a breathing mode; we find no corrugation mode. The bottom panel rules out the obvious
alternative: the spiral pattern propagates along the sound-speed characteristics, on a timescale
quite different from that of the vertical motion, so the arms are not driving it.

The origin of the vertical shear itself is less clear, and here our thermodynamic simplification
becomes an obstacle. In a real disc, vertical shear is maintained baroclinically, by a radial
temperature gradient at fixed height; a barotropic fluid has no baroclinic torque, so the shear
in our simulations must be sustained by something else. The most likely candidate is the accretion of envelope material onto the disc surfaces near
$\delta \approx \pm\pi/4$, visible in Fig.~\ref{fig: mass_flux}: this material arrives with a
specific angular momentum different from that of the disc at the same cylindrical radius, and
depositing it at the disc surface establishes a vertical gradient of $s^2\Omega$. 
% The second is
% the vertical transport of angular momentum by the spirals themselves, since the meridional
% gravitational torque is not negligible in this region (Fig.~\ref{fig: closure_nature}). 
The spiral arms, the other obvious candidate, are
disfavoured by the timescale argument above. Surface accretion would be a mechanism specific to the embedded, still-accreting phase we are simulating,
and one that has no counterpart in isolated disc models. Confirming it would require a
dedicated study.

%%%%%%%%%%%%%%%%%%%%%%%%%%%%%%%%%%%%%%%%%%%%%%%%%%%%%%%%%%%%%%%%%%%%%%%%%%%%%%%%%%%%%%%%%%%%
%%%%%%%%%%%%%%%%%%%%%%%%%%%%%%%%%%%%%%%%%%%%%%%%%%%%%%%%%%%%%%%%%%%%%%%%%%%%%%%%%%%%%%%%%%%%
\section{Discussion and conclusions}
\label{sec: discussion}

We first discuss the limitations of the present models (Sect.~\ref{subsec: limitation}), and
then summarise our conclusions (Sect.~\ref{sec: conclusion}).

%-------------------------------------------------------------------
\subsection{Limitations}
\label{subsec: limitation}

\paragraph{Thermodynamics.} The barotropic equation of state is the most consequential
simplification. It fixes the sound speed as a function of density alone, which affects the
Toomre parameter (Sect.~\ref{subsec: spirals}); it removes the baroclinic torque, which is the
standard driver of the vertical shear (Sect.~\ref{subsec: vertical_motion}); and it excludes
by construction the other thermally driven instabilities that can operate in discs, such as the
convective overstability and the zombie vortex instability
\citep[see][]{lesurHydroMagnetohydroDustGas2023}. It also carries the rescaling of the critical
densities described in Sect.~\ref{subsubsec: eos}, adopted to obtain a protostar of realistic
size. Simulations including radiative transfer will be needed to establish which of our results
survive a proper treatment of the energy budget.

\paragraph{Convergence.} As discussed in Sect.~\ref{subsec: mri}, the MRI is not resolved well
enough for the inner-disc transport to be quantitatively converged, and the higher-resolution
run differs measurably from the reference one. The partition between Maxwell and Reynolds
stresses inside $0.1~\unit{AU}$ should therefore be read as indicative.

\paragraph{The inner boundary.} The zero-gradient condition imposed on the tangential magnetic
field at $r_\mathrm{in}$ (Sect.~\ref{subsec: boundary}) generates artificial currents, and hence
some magnetic field amplification in the immediate vicinity of the boundary. We have verified
that this does not affect our results by repeating the simulation with a strongly enhanced Ohmic
resistivity in the innermost cells, which suppresses the artefact without changing either the
field structure or the evolution of the protostar; the test is described in
Appendix~\ref{app: inner_boundary}.

\paragraph{The resistivity model.} The diffusivities are read from equilibrium tables computed
for a fixed MRN grain distribution (Sect.~\ref{subsubsec: resistivities}), and the temperature
used to evaluate them in the polar cavities is set by the floor of
Sect.~\ref{subsubsec: temperature_floor}, which assumes an optically thin medium and is
therefore optimistic. A run without that floor develops a measurably different magnetic
evolution on timescales longer than our analysis window. This matters here because the entire
structure of the transport hinges on a single quantity: the radius at which the field recouples
to the gas, which separates the magnetically active inner disc from the dead zone. Anything
that moves that radius moves the boundary between the two transport regimes. Dust evolution is
one such effect: \citet{borderiesDustEvolutionChemisputtering2025} showed that the dynamical
vaporisation of grains in the innermost regions can change the resistivities substantially with
respect to the equilibrium tables used here, precisely in the range of temperatures and
densities encountered near $0.1~\unit{AU}$.

\paragraph{The azimuthal domain.} Restricting the domain to $\varphi \in [0,\pi]$
(Sect.~\ref{subsec: restart}) suppresses all odd azimuthal modes. This was
necessary to keep the protostar at the centre of the grid, but it excludes $m=1$ structures,
which can contribute to gravitational torques in massive discs and mediate the displacement of
the star from the centre of mass. The $m=2$ mode we observe is the dominant one for a
self-gravitating disc, so we do not expect this to change our conclusions qualitatively, but
the gravitational torque may be underestimated.

\paragraph{Integration time and the steady-state assumption.} The reference run covers
$\sim 64~\unit{yr}$ after the formation of the protostar, of which the analysis uses the thirty
years between $t=10$ and $40~\unit{yr}$. This is many orbital times in the inner disc, where the
analysis is most secure, but only a few at the outer edge, which is why the steady-state
description degrades there (Sect.~\ref{sec: angular_momentum_equilibrium}). The quasi-steady
phase is also interrupted by an accretion burst around $43~\unit{yr}$
(Sect.~\ref{subsec: protostar_evolution}), which we excluded from the analysis since the
steady-state assumption does not hold during it. Whether the quasi-steady regime persists as the
disc grows and the envelope depletes will require longer integrations.

\paragraph{The polar regions.} The density floors, the Alfv\'en speed limiter and the azimuthal
grid coarsening are all active in the low-density cavities above the disc
(Sect.~\ref{subsec: limiters}). The polar ejections launched from these regions should not be
taken at face value, and the magnetic braking of the envelope reported in
Sect.~\ref{sec: transport_mechanisms}, which partly operates in the same region, should be
regarded as qualitative.

\paragraph{The Hall effect.} Our model includes Ohmic and ambipolar diffusion but not the Hall
effect, which can be comparable to the other two non-ideal effects over part of the density
range relevant here \citep{marchandChemicalSolverCompute2016,tsukamotoRoleMagneticFields2023}.
Including it is a numerical challenge: the Hall term is dispersive, introducing whistler waves
whose phase speed increases as their wavelength decreases, which severely restricts the time step.
In \textsc{Idefix}, moreover, it cannot be integrated with the super-time-stepping scheme used for
the Ohmic and ambipolar terms (Sect.~\ref{subsubsec: numerical_scheme}), so that this time step
restriction applies in full, and it can only be used with the HLL solver
\citep{lesurIDEFIXVersatilePerformanceportable2023}, which suppresses the MRI at our resolution
(Sect.~\ref{subsec: mri}). It would also complicate the analysis of the effects studied here:
since the Hall effect depends on the orientation of the magnetic field with respect to the
rotation axis, the evolution differs depending on whether the two are aligned or anti-aligned
\citep{wursterRoleMagneticFields2018,tsukamotoRoleMagneticFields2023}. It may nonetheless modify
the extent of the dead zone and the transport within it
\citep{lesurThanatologyProtoplanetaryDiscs2014}, and should be included in future studies.

%-------------------------------------------------------------------
\subsection{Conclusions}
\label{sec: conclusion}

We have performed three-dimensional non-ideal MHD simulations of the collapse of a
magnetised, rotating dense core, followed continuously from the cloud scale down to the
formation of the protostar and then for $\sim 64~\unit{yr}$ beyond it, and we have used them to
analyse how angular momentum is transported in the resulting protostar--disc system. The
analysis rests on a decomposition of the azimuthally averaged angular momentum budget into
Reynolds, Maxwell and gravitational torques, each split into a radial and a meridional
component, evaluated in spherical geometry so as not to presuppose a thin disc. Our main
results are the following.

\begin{enumerate}
    \item During the quasi-steady phase that follows its formation, the angular momentum of
    the protostar stays nearly constant while its mass slowly increases, so that its specific
    angular momentum decreases: the angular momentum brought in by accretion is extracted from
    the protostar. This phase is interrupted at
    $\sim 43~\unit{yr}$ by an accretion burst, during which the protostar gains nearly a tenth
    of its mass. The results that follow describe the quiescent phase, and are robust across
    it.

    \item Accretion through the disc is vertically layered. Between $\sim 0.05$ and
    $\sim 1~\unit{AU}$, accreting and decreting layers coexist at the same cylindrical radius
    at different heights, and material is expelled from the mid-plane just outside the
    protostar even as the protostar gains mass overall.

    \item Angular momentum transport controls the accretion pattern: layers from which the stresses
    remove angular momentum accrete, layers into which they deposit it decrete. Since the meridional transport is comparable to
    the radial one, this structure is produced by a vertical redistribution of angular momentum
    that a vertically integrated, one-dimensional description cannot represent.

    \item As a consequence, a one-dimensional viscous model fitted to the net mass flux of the
    disc would require a negative equivalent $\alpha$ between $\sim 0.09$ and
    $\sim 0.35~\unit{AU}$: the vertical transport makes angular momentum accumulate in the disc
    interior, which a positive viscosity cannot do.

    \item The transport separates into three regimes, delimited by the two radii at which the
    magnetic field changes its relationship to the gas. In the envelope, where the field is coupled to the gas and
    strong, magnetic braking removes angular momentum from the infalling material. Inside $\sim 0.1~\unit{AU}$,
    where the field recouples to the gas ($\Lambda > 1$) and is wound into a strong toroidal
    component, the Maxwell torque
    dominates, with a comparable Reynolds contribution. Between the two, in the ambipolar dead
    zone ($\Lambda \ll 1$) that makes up the bulk of the disc, transport is purely hydrodynamical and
    gravitational; this dead zone itself divides into a middle disc
    ($0.1-1~\unit{AU}$), where the transport is predominantly meridional, and an outer disc
    ($1-3~\unit{AU}$), where radial and meridional transport are comparable.

    \item In the dead zone the Reynolds torque is driven by the non-axisymmetric fluctuations
    rather than by the mean flow, the former exceeding the latter by about two orders of
    magnitude. A decomposition into azimuthal modes shows that these fluctuations are dominated
    by the $m=2$ mode: the agents are the spiral arms of a disc that is an order of magnitude more
    massive than the protostar and marginally gravitationally unstable, with
    $Q_\mathrm{3D} \sim 1$ over precisely the region where the Reynolds torque dominates. 
    % The
    % disc therefore regulates its own accretion gravitationally, without magnetic help.

    \item In the inner disc, the magneto-rotational instability operates, as shown by the
    butterfly pattern of the azimuthally averaged field. It both saturates the field, which
    reaches $3-10~\unit{kG}$ in the protostar region instead of being amplified indefinitely by
    the shear, and feeds
    the velocity fluctuations that make the Reynolds torque comparable to the Maxwell torque
    there. The instability is, however, not numerically converged.

    \item A systematic mean vertical motion is present in the disc, anti-correlated with the
    vertical shear and propagating outwards well below the sound speed, with the properties of
    a vertical shear instability breathing mode. It contributes negligibly to the angular
    momentum budget, but may matter for dust settling and mixing.
\end{enumerate}

The broader implication is that no single mechanism transports angular momentum in a young
protostar--disc system. The magnetic field controls the two ends of the problem --- the
envelope and the innermost disc --- while the bulk of the disc, magnetically decoupled by
ambipolar diffusion, relies on hydrodynamical and gravitational stresses. The boundaries between these
regimes are not set by the disc structure but by the microphysics of the gas--dust mixture,
through the radius at which the field recouples to the gas. Improving the resistivity model,
and in particular accounting for the dynamical vaporisation of dust grains
\citep{borderiesDustEvolutionChemisputtering2025}, is therefore not a refinement of a detail: it
determines where one transport regime gives way to another.

These results also bear on how young discs are modelled.
One-dimensional viscous models describe the transport by a single $\alpha$
parameter \citep{shakuraBlackHolesBinary1973}; the layered accretion and the negative equivalent
$\alpha$ found here show that such a description cannot capture the transport in a young, thick
and embedded disc, in which the vertical redistribution of angular momentum matters as much as
the radial one.

The natural extensions of this work follow from the limitations of
Sect.~\ref{subsec: limitation}: a proper treatment of the energy budget including radiative
transfer, a higher resolution in the inner disc, and a full azimuthal domain. Finally, our analysis has taken the magnetic field of the inner disc as given.
The mechanism that amplifies it and sustains its poloidal component against dissipation, and
hence the Maxwell torque itself, will be the subject of a forthcoming paper
(Borderies et al., in prep.).

\section*{Acknowledgements}
We thank Kengo Tomida, Shinsuke Takasao, Geoffroy Lesur, Fran\c{c}ois Rincon and Mario Flock
for fruitful discussions. This work has received funding from the French Agence Nationale de la Recherche (ANR) through
the projects PROMETHEE (ANR-22-CE31-0020) and DISKBUILD (ANR-20-CE49-0006). This work was
granted access to the HPC resources of IDRIS (Jean Zay) under the allocation
AD010417023 made by GENCI. We gratefully acknowledge the support of the Centre Blaise
Pascal's IT test platform at ENS de Lyon (Lyon, France) for GPU computing facilities. The platform
operates the SIDUS solution \citep{quemenerSIDUSSolutionExtreme2013} developed by Emmanuel
Quemener. The text of this manuscript was drafted and edited with the assistance of the large
language model Claude (Anthropic). All simulations, analyses and figures were produced by the authors without generative AI, all references were
checked by the authors, and the authors reviewed the entire text and take full responsibility for
its content.

\section*{Data availability}
The simulation data and the analysis scripts underlying this article will be shared on
reasonable request to the corresponding author. The \textsc{Idefix} code is publicly available
at \url{https://github.com/idefix-code/idefix}.

\bibliographystyle{aa} % style aa.bst
\bibliography{biblio} % your references

\begin{thebibliography}{56}
\expandafter\ifx\csname natexlab\endcsname\relax\def\natexlab#1{#1}\fi

\bibitem[{Ahmad {et~al.}(2023)Ahmad, Gonz{\'a}lez, Hennebelle, \& Commer{\c c}on}]{ahmadBirthEarlyEvolution2023}
Ahmad, A., Gonz{\'a}lez, M., Hennebelle, P., \& Commer{\c c}on, B. 2023, A\&A, 680, A23

\bibitem[{Ahmad {et~al.}(2024)Ahmad, Gonz{\'a}lez, Hennebelle, \& Commer{\c c}on}]{ahmadFormationLowmassProtostars2024}
Ahmad, A., Gonz{\'a}lez, M., Hennebelle, P., \& Commer{\c c}on, B. 2024, A\&A, 687, A90

\bibitem[{Ahmad {et~al.}(2025)Ahmad, Gonz{\'a}lez, Hennebelle, Lebreuilly, \& Commer{\c c}on}]{ahmadBirthMagnetizedLowmass2025}
Ahmad, A., Gonz{\'a}lez, M., Hennebelle, P., Lebreuilly, U., \& Commer{\c c}on, B. 2025, A\&A, 696, A238

\bibitem[{Ahmad {et~al.}(2026)Ahmad, Commer{\c c}on, Lynch, Lovascio, Charnoz, Marschall, \& Morbidelli}]{ahmadNumericalInsightsDisk2026}
Ahmad, A.~A., Commer{\c c}on, B., Lynch, E., {et~al.} 2026, Astronomy and Astrophysics, 707, A97

\bibitem[{Balbus \& Hawley(1991)}]{balbusPowerfulLocalShear1991}
Balbus, S.~A. \& Hawley, J.~F. 1991, The Astrophysical Journal, 376, 214

\bibitem[{Bhandare {et~al.}(2020)Bhandare, Kuiper, Henning, Fendt, Flock, \& Marleau}]{bhandareBirthConvectiveLowmass2020}
Bhandare, A., Kuiper, R., Henning, T., {et~al.} 2020, A\&A, 638, A86

\bibitem[{Bhandare {et~al.}(2018)Bhandare, Kuiper, Henning, Fendt, Marleau, \& K{\"o}lligan}]{bhandareFirstCoreProperties2018}
Bhandare, A., Kuiper, R., Henning, T., {et~al.} 2018, Astronomy and Astrophysics, 618, A95

\bibitem[{Borderies {et~al.}(2025)Borderies, Commer{\c c}on, \& Bourdon}]{borderiesDustEvolutionChemisputtering2025}
Borderies, A., Commer{\c c}on, B., \& Bourdon, B. 2025, A\&A, 694, A89

\bibitem[{Commer{\c c}on {et~al.}(2024)Commer{\c c}on, Lovascio, Lynch, \& Ragusa}]{commerconDiscsAreBorn2024}
Commer{\c c}on, B., Lovascio, F., Lynch, E., \& Ragusa, E. 2024, Astronomy and Astrophysics, 689, L9

\bibitem[{Cossins {et~al.}(2009)Cossins, Lodato, \& Clarke}]{cossinsCharacterizingGravitationalInstability2009}
Cossins, P., Lodato, G., \& Clarke, C.~J. 2009, Monthly Notices of the Royal Astronomical Society, 393, 1157

\bibitem[{Crida {et~al.}(2025)Crida, Baruteau, Gonzalez, Masset, Segretain, Griveaud, M{\'e}heut, \& Lega}]{cridaReflexInstabilityExponential2025}
Crida, A., Baruteau, C., Gonzalez, J.-F., {et~al.} 2025, The Open Journal of Astrophysics, 8, 115

\bibitem[{Durisen {et~al.}(1986)Durisen, Gingold, Tohline, \& Boss}]{durisenDynamicFissionInstabilities1986}
Durisen, R.~H., Gingold, R.~A., Tohline, J.~E., \& Boss, A.~P. 1986, The Astrophysical Journal, 305, 281

\bibitem[{Einfeldt {et~al.}(1991)Einfeldt, Roe, Munz, \& Sjogreen}]{einfeldtGodunovTypeMethodsLow1991}
Einfeldt, B., Roe, P.~L., Munz, C.~D., \& Sjogreen, B. 1991, Journal of Computational Physics, 92, 273

\bibitem[{Evans \& Hawley(1988)}]{evansSimulationMagnetohydrodynamicFlows1988}
Evans, C.~R. \& Hawley, J.~F. 1988, The Astrophysical Journal, 332, 659

\bibitem[{Flock {et~al.}(2010)Flock, Dzyurkevich, Klahr, \& Mignone}]{flockHighorderGodunovSchemes2010}
Flock, M., Dzyurkevich, N., Klahr, H., \& Mignone, A. 2010, Astronomy and Astrophysics, 516, A26

\bibitem[{Fromang(2013)}]{fromangMRIdrivenAngularMomentum2013}
Fromang, S. 2013, in {{EAS Publications Series}}, Vol.~62 (eprint: arXiv:1305.3416: EDP), 95--142

\bibitem[{Galli \& Shu(1993)}]{galliCollapseMagnetizedMolecular1993}
Galli, D. \& Shu, F.~H. 1993, The Astrophysical Journal, 417, 243

\bibitem[{Harten {et~al.}(1983)Harten, Lax, \& van Leer}]{hartenUpstreamDifferencingGodunovType1983}
Harten, A., Lax, P.~D., \& van Leer, B. 1983, SIAM Rev., 25, 35

\bibitem[{Hawley {et~al.}(1995)Hawley, Gammie, \& Balbus}]{hawleyLocalThreedimensionalMagnetohydrodynamic1995}
Hawley, J.~F., Gammie, C.~F., \& Balbus, S.~A. 1995, The Astrophysical Journal, 440, 742

\bibitem[{Hennebelle(2013)}]{hennebelleTransportInfluenceAngular2013}
Hennebelle, P. 2013, in {{EAS Publications Series}}, Vol.~62 (EDP), 67--94

\bibitem[{Hennebelle \& Fromang(2008)}]{hennebelleMagneticProcessesCollapsing2008}
Hennebelle, P. \& Fromang, S. 2008, A\&A, 477, 9

\bibitem[{Kratter \& Lodato(2016)}]{kratterGravitationalInstabilitiesCircumstellar2016}
Kratter, K. \& Lodato, G. 2016, Annual Review of Astronomy and Astrophysics, 54, 271

\bibitem[{Laughlin {et~al.}(1997)Laughlin, Korchagin, \& Adams}]{laughlinSpiralModeSaturation1997}
Laughlin, G., Korchagin, V., \& Adams, F.~C. 1997, The Astrophysical Journal, 477, 410

\bibitem[{Lesur {et~al.}(2023{\natexlab{a}})Lesur, Flock, Ercolano, Lin, Yang, Barranco, {Benitez-Llambay}, Goodman, Johansen, Klahr, Laibe, Lyra, Marcus, Nelson, Squire, Simon, Turner, Umurhan, \& Youdin}]{lesurHydroMagnetohydroDustGas2023}
Lesur, G., Flock, M., Ercolano, B., {et~al.} 2023{\natexlab{a}}, Hydro-, {{Magnetohydro}}-, and {{Dust-Gas Dynamics}} of {{Protoplanetary Disks}}

\bibitem[{Lesur {et~al.}(2014)Lesur, Kunz, \& Fromang}]{lesurThanatologyProtoplanetaryDiscs2014}
Lesur, G., Kunz, M.~W., \& Fromang, S. 2014, Astronomy and Astrophysics, 566, A56

\bibitem[{Lesur {et~al.}(2023{\natexlab{b}})Lesur, Baghdadi, {Wafflard-Fernandez}, Mauxion, Robert, \& {Van den Bossche}}]{lesurIDEFIXVersatilePerformanceportable2023}
Lesur, G. R.~J., Baghdadi, S., {Wafflard-Fernandez}, G., {et~al.} 2023{\natexlab{b}}, A\&A, 677, A9

\bibitem[{Machida \& Basu(2019)}]{machidaFirstTwoThousand2019}
Machida, M.~N. \& Basu, S. 2019, The Astrophysical Journal, 876, 149

\bibitem[{Marchand {et~al.}(2016)Marchand, Masson, Chabrier, Hennebelle, Commer{\c c}on, \& Vaytet}]{marchandChemicalSolverCompute2016}
Marchand, P., Masson, J., Chabrier, G., {et~al.} 2016, A\&A, 592, A18

\bibitem[{Maury {et~al.}(2019)Maury, Andr{\'e}, Testi, Maret, Belloche, Hennebelle, Cabrit, Codella, Gueth, Podio, Anderl, Bacmann, Bontemps, Gaudel, Ladjelate, Lef{\`e}vre, Tabone, \& Lefloch}]{mauryCharacterizingYoungProtostellar2019}
Maury, A.~J., Andr{\'e}, {\relax Ph}., Testi, L., {et~al.} 2019, Astronomy and Astrophysics, 621, A76

\bibitem[{Mauxion(2024)}]{mauxionInteractionDisquesProtoplanetaires2024}
Mauxion, J. 2024, These de doctorat, Universit\'e Grenoble Alpes

\bibitem[{Mauxion {et~al.}(2024)Mauxion, Lesur, \& Maret}]{mauxionModelingSecularEvolution2024}
Mauxion, J., Lesur, G., \& Maret, S. 2024, Astronomy and Astrophysics, 686, A253

\bibitem[{Mellon \& Li(2008)}]{mellonMagneticBrakingProtostellar2008}
Mellon, R.~R. \& Li, Z.-Y. 2008, The Astrophysical Journal, 681, 1356

\bibitem[{Meyer {et~al.}(2014)Meyer, Balsara, \& Aslam}]{meyerStabilizedRungeKutta2014}
Meyer, C.~D., Balsara, D.~S., \& Aslam, T.~D. 2014, Journal of Computational Physics, 257, 594

\bibitem[{Miyoshi \& Kusano(2005)}]{miyoshiMultistateHLLApproximate2005}
Miyoshi, T. \& Kusano, K. 2005, Journal of Computational Physics, 208, 315

\bibitem[{Mouschovias \& Paleologou(1980)}]{mouschoviasMagneticBrakingAligned1980}
Mouschovias, T.~{\relax Ch}. \& Paleologou, E.~V. 1980, The Astrophysical Journal, 237, 877

\bibitem[{Mouschovias \& Spitzer(1976)}]{mouschoviasNoteCollapseMagnetic1976}
Mouschovias, T.~{\relax Ch}. \& Spitzer, Jr., L. 1976, ApJ, 210, 326

\bibitem[{Nelson {et~al.}(2013)Nelson, Gressel, \& Umurhan}]{nelsonLinearNonlinearEvolution2013}
Nelson, R.~P., Gressel, O., \& Umurhan, O.~M. 2013, Monthly Notices of the Royal Astronomical Society, 435, 2610

\bibitem[{Ogilvie {et~al.}(2025)Ogilvie, Latter, \& Lesur}]{ogilvieVerticalShearInstability2025}
Ogilvie, G.~I., Latter, H.~N., \& Lesur, G. 2025, Monthly Notices of the Royal Astronomical Society, 537, 3349

\bibitem[{Ohashi {et~al.}(2014)Ohashi, Saigo, Aso, Aikawa, Koyamatsu, Machida, Saito, Takahashi, Takakuwa, Tomida, Tomisaka, \& Yen}]{ohashiFormationKeplerianDisk2014}
Ohashi, N., Saigo, K., Aso, Y., {et~al.} 2014, The Astrophysical Journal, 796, 131

\bibitem[{Pringle(1981)}]{pringleAccretionDiscsAstrophysics1981}
Pringle, J.~E. 1981, Annual Review of Astronomy and Astrophysics, 19, 137

\bibitem[{Quemener \& Corvellec(2013)}]{quemenerSIDUSSolutionExtreme2013}
Quemener, E. \& Corvellec, M. 2013, Linux J., 2013, 3:3

\bibitem[{Saad(2003)}]{saadIterativeMethodsSparse2003}
Saad, Y. 2003, Iterative {{Methods}} for {{Sparse Linear Systems}} ({Society for Industrial and Applied Mathematics})

\bibitem[{Shakura \& Sunyaev(1973)}]{shakuraBlackHolesBinary1973}
Shakura, N.~I. \& Sunyaev, R.~A. 1973, Astronomy and Astrophysics, 24, 337

\bibitem[{Tobin {et~al.}(2012)Tobin, Hartmann, Chiang, Wilner, Looney, Loinard, Calvet, \& D'Alessio}]{tobin02solarmassProtostarKeplerian2012}
Tobin, J.~J., Hartmann, L., Chiang, H.-F., {et~al.} 2012, Nature, 492, 83

\bibitem[{Tobin {et~al.}(2020)Tobin, Sheehan, Megeath, {D{\'i}az-Rodr{\'i}guez}, Offner, Murillo, {van 't Hoff}, {van Dishoeck}, Osorio, Anglada, Furlan, Stutz, Reynolds, Karnath, Fischer, Persson, Looney, Li, Stephens, Chandler, Cox, Dunham, Tychoniec, Kama, Kratter, Kounkel, Mazur, Maud, Patel, Perez, Sadavoy, {Segura-Cox}, Sharma, Stephenson, Watson, \& Wyrowski}]{tobinVLAALMANascent2020}
Tobin, J.~J., Sheehan, P.~D., Megeath, S.~T., {et~al.} 2020, The Astrophysical Journal, 890, 130

\bibitem[{Tohline {et~al.}(1985)Tohline, Durisen, \& McCollough}]{tohlineLinearNonlinearDynamic1985}
Tohline, J.~E., Durisen, R.~H., \& McCollough, M. 1985, The Astrophysical Journal, 298, 220

\bibitem[{Toomre(1964)}]{toomreGravitationalStabilityDisk1964}
Toomre, A. 1964, The Astrophysical Journal, 139, 1217

\bibitem[{Trott {et~al.}(2022)Trott, {Lebrun-Grandie}, Arndt, Ciesko, Dang, Ellingwood, Gayatri, Harvey, Hollman, Ibanez, Liber, Madsen, Miles, Poliakoff, Powell, Rajamanickam, Simberg, Sunderland, Turcksin, \& Wilke}]{trottKokkos3Programming2022}
Trott, C.~R., {Lebrun-Grandie}, D., Arndt, D., {et~al.} 2022, IEEE Transactions on Parallel and Distributed Systems, 33, 805

\bibitem[{Tsukamoto {et~al.}(2023)Tsukamoto, Maury, Commercon, Alves, Cox, Sakai, Ray, Zhao, \& Machida}]{tsukamotoRoleMagneticFields2023}
Tsukamoto, Y., Maury, A., Commercon, B., {et~al.} 2023, The {{Role}} of {{Magnetic Fields}} in the {{Formation}} of {{Protostars}}, {{Disks}}, and {{Outflows}}

\bibitem[{Vaidya {et~al.}(2017)Vaidya, Prasad, Mignone, Sharma, \& Rickler}]{vaidyaScalableExplicitImplementation2017}
Vaidya, B., Prasad, D., Mignone, A., Sharma, P., \& Rickler, L. 2017, Monthly Notices of the Royal Astronomical Society, 472, 3147

\bibitem[{{van Leer}(1979)}]{vanleerUltimateConservativeDifference1979}
{van Leer}, B. 1979, Journal of Computational Physics, 32, 101

\bibitem[{Vaytet {et~al.}(2018)Vaytet, Commer{\c c}on, Masson, Gonz{\'a}lez, \& Chabrier}]{vaytetProtostellarBirthAmbipolar2018}
Vaytet, N., Commer{\c c}on, B., Masson, J., Gonz{\'a}lez, M., \& Chabrier, G. 2018, A\&A, 615, A5

\bibitem[{Weast(1981)}]{weastCRCHandbookChemistry1981}
Weast, R. 1981, {{CRC Handbook}} of {{Chemistry}} and {{Physics}} (Boca Raton: CRC)

\bibitem[{Wurster \& Li(2018)}]{wursterRoleMagneticFields2018}
Wurster, J. \& Li, Z.-Y. 2018, Frontiers in Astronomy and Space Sciences, 5, 39

\bibitem[{Zhao {et~al.}(2016)Zhao, Caselli, Li, Krasnopolsky, Shang, \& Nakamura}]{zhaoProtostellarDiscFormation2016}
Zhao, B., Caselli, P., Li, Z.-Y., {et~al.} 2016, MNRAS, 460, 2050

\bibitem[{Zhu \& Stone(2018)}]{zhuGlobalEvolutionAccretion2018}
Zhu, Z. \& Stone, J.~M. 2018, The Astrophysical Journal, 857, 34

\end{thebibliography}

\begin{appendix}

\section{Restart procedure}
\label{app: restart}

\subsection{Radial restart}
\label{subsubsec: radial_restart}

\begin{figure*}[tb]
    \centering
    \includegraphics[width=\textwidth]{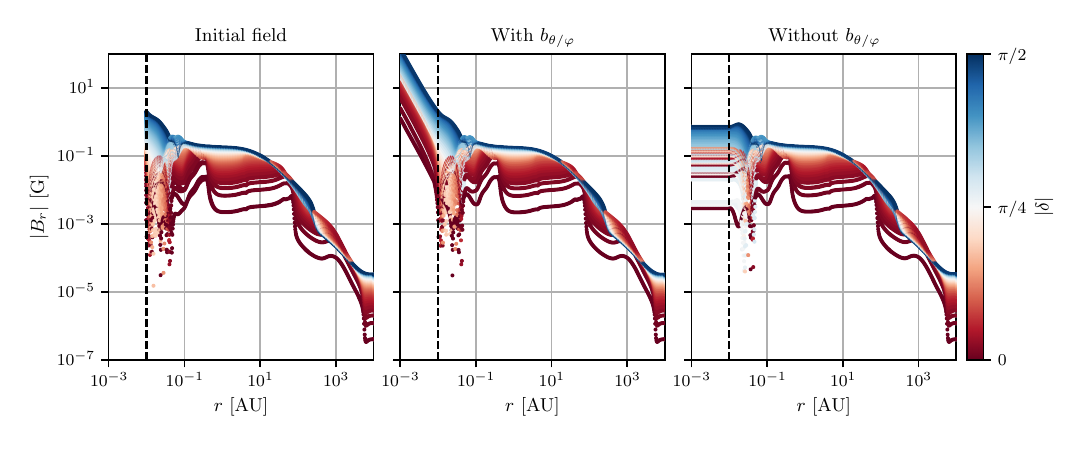}
    \caption{Radial magnetic field during a radial restart, from
    $r_\mathrm{in,old} = 10^{-2}~\unit{AU}$ (vertical dashed line) to
    $r_\mathrm{in,new} = 10^{-3}~\unit{AU}$. Colours show
    $|\delta| = |\pi/2 - \theta|$, the angle to the equatorial plane.
    \textit{Left:} field before the restart. \textit{Middle:} the continuity-preserving
    extrapolation, which amplifies $B_r$ by several orders of magnitude near the centre through
    the singular $b_\theta$ and $b_\varphi$ terms. \textit{Right:} the extrapolation adopted
    here, Eq.~\eqref{eq: extrapolation}, which removes the singularity while preserving the
    inner field structure.}
    \label{fig: extrapolation}
\end{figure*}

This appendix details the restarts summarised in Sect.~\ref{subsec: restart}, in which the
extent of the domain --- but not its relative resolution --- is changed.

The hyperbolic time step scales with the innermost cell size, and hence with
$r_\mathrm{in}$. Starting the simulation with $r_\mathrm{in} = 10^{-3}~\unit{AU}$ would impose
the protostellar time step from the very beginning, long before any structure exists at that
scale. We therefore start with $r_\mathrm{in} = 1~\unit{AU}$ and reduce it by one decade at a
time, following the characteristic scale of the collapse, down to
$10^{-3}~\unit{AU}$.

Two criteria decide when the inner radius can be reduced. First, the density near
$r_\mathrm{in}$ must be nearly uniform and close to the mean density $\rho_\mathrm{in}$ inside
$r_\mathrm{in}$, which is obtained by integrating the mass flux through the inner boundary; a
significant density contrast means that the collapse is already developing at that scale.
Second, the Jeans length associated with $\rho_\mathrm{in}$ must remain larger than
$r_\mathrm{in}$, so that no unresolved collapsing structure is hidden inside the inner
boundary. In practice both criteria are met with a wide margin at each restart, the density
contrast staying at the percent level. Three reductions then fall naturally at the
successive stages of the collapse: the first is performed as the first Larson core is about to
form, the second at the onset of the second collapse, and the third, which brings the grid to
its final inner radius of $10^{-3}~\unit{AU}$, as the second Larson core is about to form.

The solution must then be extrapolated inwards. The density is set to $\rho_\mathrm{in}$ in the
new cells, and the velocity is extrapolated linearly so that it vanishes at the centre:
\begin{equation}
    \rho_\mathrm{extra}(\bm{r}) = \rho_\mathrm{in}, \qquad
    \bm{v}_\mathrm{extra}(\bm{r}) = \bm{v}(r_\mathrm{in,old},\theta,\varphi)\,
    \frac{r}{r_\mathrm{in,old}} .
\end{equation}
The magnetic field is more delicate, since the solenoidal constraint must be preserved. We
extrapolate the vector potential instead, requiring $B_\theta$ and $B_\varphi$ to remain
radially uniform and continuous inside $r_\mathrm{in,old}$. This construction leaves an
unphysical $B_r \propto r^{-2}$ singularity at the centre, which we remove by adopting
\begin{equation}
    \begin{cases}
        A_{r,\mathrm{new}} = A_{r,\mathrm{old}}(r_\mathrm{in,old},\theta,\varphi)\,
        \dfrac{r}{r_\mathrm{in,old}}, \\[6pt]
        A_{\theta,\mathrm{new}} = a_\theta \dfrac{r}{r_\mathrm{in,old}}, \qquad
        A_{\varphi,\mathrm{new}} = a_\varphi \dfrac{r}{r_\mathrm{in,old}},
    \end{cases}
    \label{eq: extrapolation}
\end{equation}
for $r < r_\mathrm{in,old}$, and
$A_{\theta,\mathrm{new}} = A_{\theta,\mathrm{old}} - b_\theta\, r_\mathrm{in,old}/r$
(and similarly for $A_\varphi$) outside, with
\begin{equation}
    \begin{cases}
        a_\theta = \frac{1}{2}\partial_r\left(rA_\theta\right)(r_\mathrm{in,old},\theta,\varphi),
        \quad b_\theta = A_\theta(r_\mathrm{in,old},\theta,\varphi) - a_\theta,\\[4pt]
        a_\varphi = \frac{1}{2}\partial_r\left(rA_\varphi\right)(r_\mathrm{in,old},\theta,\varphi),
        \quad b_\varphi = A_\varphi(r_\mathrm{in,old},\theta,\varphi) - a_\varphi.
    \end{cases}
\end{equation}
With this choice $B_\theta$ and $B_\varphi$ are preserved throughout the grid and $B_r$ is not
amplified near the centre; the residual modification of $B_r$ decreases as $r^{-2}$ and leaves
the inner field structure unchanged (Fig.~\ref{fig: extrapolation}).

\subsection{Polar restart}
\label{subsubsec: polar_restart}
A spherical grid is only efficient if the protostar forms at its centre. The first Larson core
\citep{bhandareFirstCoreProperties2018} is turbulent and accretes asymmetrically
\citep{commerconDiscsAreBorn2024,ahmadNumericalInsightsDisk2026}, and the non-symmetric structures developing during the first-core phase can displace the centre of mass through
round-off errors and asymmetric interactions with the boundaries, and produce an off-centre
protostar --- which not only wastes the inner resolution but also disturbs the velocity and
magnetic fields near the inner boundary. We therefore run the collapse and the first-core
evolution on a half-domain, $\theta \in [0,\pi/2]$ with $N_\theta = 64$ cells, which enforces
equatorial symmetry, and restart on the full sphere just before the second collapse. The lower
hemisphere is filled by symmetry,
\begin{equation}
    \begin{cases}
        \rho(r,\pi-\theta,\varphi) = \rho(r,\theta,\varphi),\\
        X_r(r,\pi-\theta,\varphi) = X_r(r,\theta,\varphi),\\
        X_\theta(r,\pi-\theta,\varphi) = -X_\theta(r,\theta,\varphi),\\
        X_\varphi(r,\pi-\theta,\varphi) = X_\varphi(r,\theta,\varphi),
    \end{cases}
\end{equation}
with $\bm{X} = \bm{v}$ or $\bm{A}$. Note that the symmetry relations are those of a vector,
not of the pseudo-vector $\bm{B}$, precisely because the field is evolved through its
potential.

\subsection{Azimuthal restart}
\label{subsubsec: azimuthal_restart}

Three-dimensional simulations are considerably more expensive, and no non-axisymmetric
structure is expected to play a significant role during the first collapse, such structures developing once the first core and its disc have
formed \citep{commerconDiscsAreBorn2024,ahmadNumericalInsightsDisk2026}. We therefore compute that stage in
axisymmetry and restart on a three-dimensional grid shortly before the formation of the first
Larson core. The axisymmetric fields are simply replicated in $\varphi$, with a density
perturbation added to seed non-axisymmetric structures:
\begin{equation}
    \rho_\mathrm{3D}(r,\theta,\varphi) = \rho_\mathrm{2D}(r,\theta)
    \left[1 + \epsilon \sin(2\varphi)\right],
\end{equation}
the velocity components being copied unchanged. We force an $m=2$ mode, the most unstable one
\citep{tohlineLinearNonlinearDynamic1985,durisenDynamicFissionInstabilities1986}, with
$\epsilon$ between $10^{-2}$ and $10^{-1}$; the precise value only sets the time needed for the
structures to develop and does not affect the results. For the vector potential, only
$A_\varphi$ is defined in two dimensions; we use the gauge freedom to set $A_r = 0$ and obtain
$A_\theta$ by integrating $\frac{1}{r}\partial_r(rA_\theta) = B_\varphi$.

We initially used a full azimuthal domain, $\varphi \in [0,2\pi]$, but found that an $m=1$ mode
systematically develops and ejects the protostar from the centre of the grid, even when the
seed perturbation contains no odd mode. Since the centre of mass itself becomes off-centre when
this happens, we attribute it to the interaction of the protostar with the inner boundary: once
a small $m=1$ mode appears, the net force exerted by the inner boundary no longer vanishes and
amplifies it. This may be related to the reflex instability described by
\citet{cridaReflexInstabilityExponential2025}, in which the reflex motion of the star excites the
eccentricity of the disc, whose $m=1$ density perturbation in turn pulls the star further away;
confirming this would require a dedicated study. We therefore restrict the domain to $\varphi \in [0,\pi]$, which suppresses all
odd modes and keeps the protostar centred, at the cost of filtering out $m=1$ physics.

\subsection{Increased-resolution restart}
\label{subsubsec: resolution_restart}

Finally, a restart at higher resolution is used for the convergence run of
Table~\ref{tab: simulations}. Each parent cell is subdivided by an integer factor in each
refined direction (here $\theta$ and $\varphi$); cell-centred variables are transferred by zeroth-order injection, and the magnetic
field is refined through its vector potential so that the solenoidal constraint is preserved
exactly.

\section{The HLL Riemann solver}
\label{app: hll}

\begin{figure*}[t]
    \centering
    \includegraphics[width=1\textwidth]{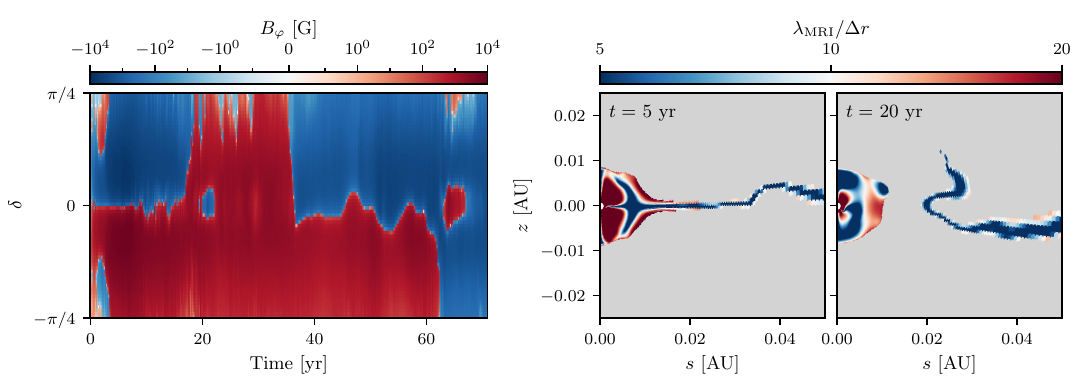}
    \caption{Magneto-rotational instability in the 1MsunHLL run. \textit{Left:} butterfly
    diagram of the azimuthally averaged toroidal field $B_\varphi$ at $r=0.015~\unit{AU}$, as a
    function of time and of the angle $\delta$ to the equatorial plane, to be compared with
    Fig.~\ref{fig: butterfly_x2}. \textit{Right:} number of cells per wavelength of the most
    unstable MRI mode at $t=5$ and $20~\unit{yr}$, as in Fig.~\ref{fig: mri_resolution}; regions
    where $\lambda_\mathrm{MRI}$ exceeds the local disc height are shown in grey.}
    \label{fig: hll_mri}
\end{figure*}

\begin{figure*}[t]
    \centering
    \includegraphics[width=1\textwidth]{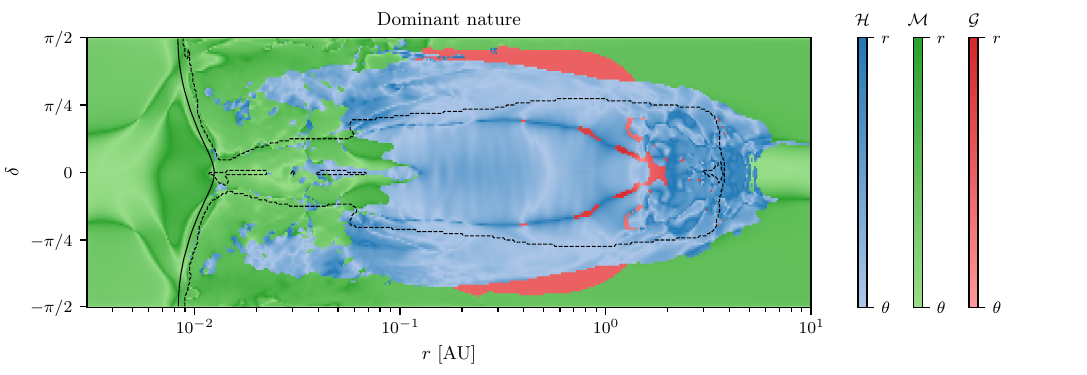}
    \caption{Dominant torque at each location in the 1MsunHLL run, averaged in time between
    $t=10~\unit{yr}$ and $t=40~\unit{yr}$, as in the top panel of
    Fig.~\ref{fig: closure_nature}.}
    \label{fig: hll_nature}
\end{figure*}

All the results presented in the main text use the HLLD Riemann solver
\citep{miyoshiMultistateHLLApproximate2005}. For comparison, we also ran the reference setup
with the more diffusive HLL solver
\citep{hartenUpstreamDifferencingGodunovType1983,einfeldtGodunovTypeMethodsLow1991}, all other
parameters being identical (Table~\ref{tab: simulations}). This appendix collects the
differences.

The essential one is the fate of the magneto-rotational instability in the inner disc.
Figure~\ref{fig: hll_mri} shows that, at our resolution, the numerical dissipation of the HLL
solver is large enough to suppress it. The region in which the most unstable mode fits within
the disc is thinner than with HLLD (Fig.~\ref{fig: mri_resolution}), and the mode is resolved by
fewer than ten cells over most of it. No butterfly pattern develops: the toroidal field keeps the
ordered configuration that the HLLD run displays before the onset of the instability
(Fig.~\ref{fig: butterfly_x2}), negative above the mid-plane and positive below it. It changes
sign only once over the run, in the northern hemisphere between $\sim 20$ and
$\sim 35~\unit{yr}$, without the quasi-periodic, migrating reversals of the HLLD run. Without
MRI turbulence to saturate it, the field is also stronger than in the HLLD run, by a factor of
$\sim 2$ overall, although of the same order of magnitude.

The consequences for the transport are shown in Fig.~\ref{fig: hll_nature}. Inside
$\sim 0.1~\unit{AU}$, the Maxwell torque dominates almost the whole inner disc, whereas in the
HLLD run it shares this region with the Reynolds torque (Fig.~\ref{fig: closure_nature}). This is
consistent with the contribution of MRI turbulence to the velocity fluctuations in the HLLD run
(Sect.~\ref{subsec: mri}). Outside the recoupling radius, where the field is decoupled and the
transport is hydrodynamical and gravitational, the two runs show the same regions: the Reynolds
torque dominates the disc, predominantly through its meridional component in the middle disc,
and the Maxwell torque dominates the envelope.

\section{The inner boundary condition}
\label{app: inner_boundary}

\begin{figure*}[t]
    \centering
    \includegraphics[width=\textwidth]{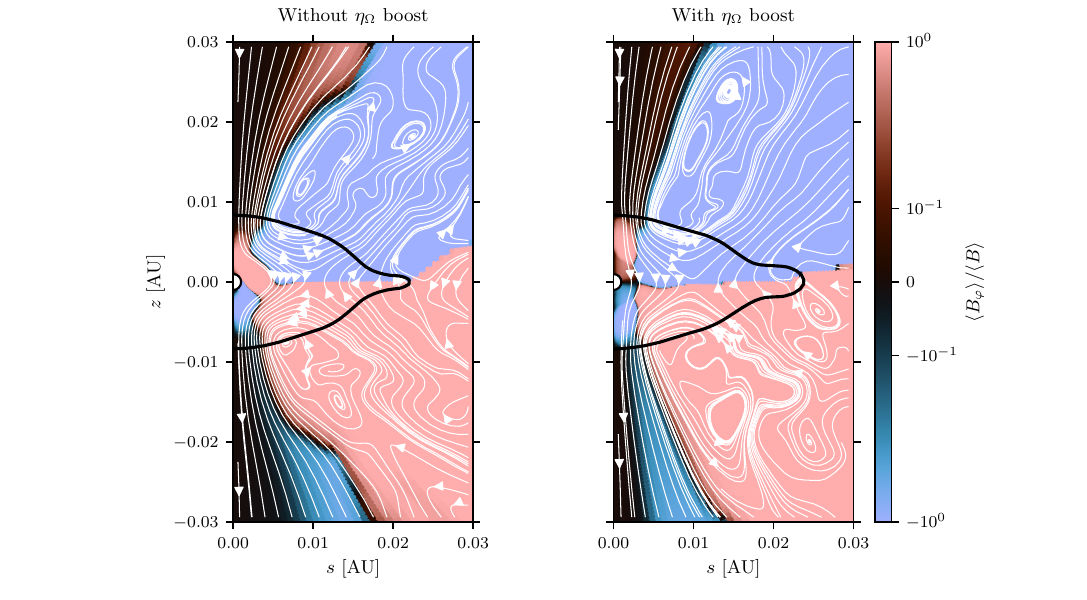}
    \caption{Magnetic field structure at the protostellar scale, without (\textit{left}) and
    with (\textit{right}) the enhanced Ohmic resistivity of Eq.~\eqref{eq: boosted_resistivity}.
    Colours show the azimuthally averaged ratio $\averagePhi{B_\varphi}/\averagePhi{B}$ and
    white lines the poloidal field. The black contour marks the protostar surface.}
    \label{fig: boost_ohm}
% \end{figure*}

% \begin{figure*}[t]
    \centering
    \includegraphics[width=\textwidth]{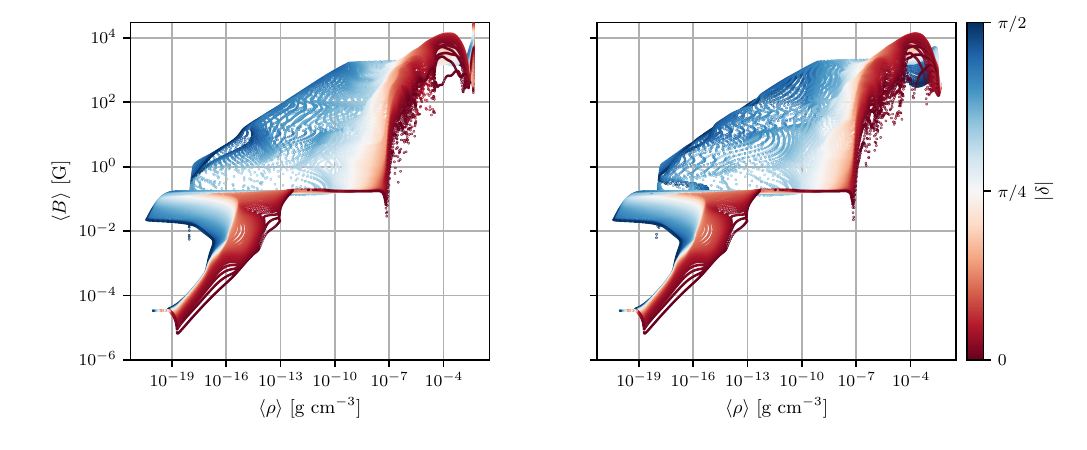}
    \caption{Azimuthally averaged magnetic field strength as a function of density, without
    (\textit{left}) and with (\textit{right}) the enhanced Ohmic resistivity. Colours show
    $|\delta| = |\pi/2-\theta|$. The narrow spike reaching $\sim 2\times 10^{4}~\unit{G}$ at the
    highest densities in the left panel is the boundary artefact; it is absent on the right.}
    \label{fig: boost_ohm_scatter}
\end{figure*}

The magnetic boundary condition at the inner radius imposes a vanishing gradient on $B_\theta$
and $B_\varphi$ (Sect.~\ref{subsec: boundary}). This is not a physical condition: it forces
currents to close through the boundary, and those artificial currents amplify the field in the
innermost cells. Since the Maxwell torque dominates the transport in exactly this part of the
domain (Sect.~\ref{sec: transport_mechanisms}), it is necessary to establish that the
amplification is confined to the boundary layer and does not contaminate the inner disc.

The cleanest way to remove the artefact without altering the physics elsewhere is to decouple
the field from the gas in the immediate vicinity of the boundary, so that no current can be
sustained there. We therefore repeated the simulation with an Ohmic resistivity enhanced close
to $r_\mathrm{in}$,
\begin{equation}
    \eta_\Omega = \frac{1}{2}\left(1-\tanh\left(\frac{r-r_\Omega}{\delta r_\Omega}\right)\right)
    \eta_{\Omega,\mathrm{boost}} + \eta_{\Omega,0},
    \label{eq: boosted_resistivity}
\end{equation}
where $\eta_{\Omega,0}$ is the resistivity read from the tables of
\citet{marchandChemicalSolverCompute2016}, $r_\Omega = 3\,r_\mathrm{in}$ is the transition
radius and $\delta r_\Omega = r_\Omega/30$ its width, the $\tanh$ ensuring a smooth transition.
We use $\eta_{\Omega,\mathrm{boost}} = 10^{10}\,\eta_{\Omega,0}$. Since the physical resistivity
is very small at the centre of the protostar, this amounts to an essentially complete
decoupling over the first few cells.

This experiment was carried out with the HLL solver rather than with HLLD. The artefact it
addresses is a property of the boundary condition and of the innermost cells, not of the
Riemann solver: the magnetic structure at the protostar--disc interface is qualitatively the
same in the two runs, with a dominant toroidal component of opposite sign on either side of the
mid-plane and a comparable field strength, and both show the same amplification against the
boundary. The principal difference between the solvers, the presence of the MRI
(Sect.~\ref{subsec: mri}), develops at radii several times larger than the region in which the
resistivity is enhanced here, and so does not bear on what is being tested. We have therefore
not repeated the experiment with HLLD.

Figures~\ref{fig: boost_ohm} and~\ref{fig: boost_ohm_scatter} compare the two runs. The
enhanced resistivity does what it was designed to do: in the $B-\rho$ relation, the narrow
spike at the highest densities --- the signature of the field being amplified against the
boundary --- disappears. Everything else is unchanged. The field retains the same topology at
the protostellar scale, with the toroidal component dominant and antisymmetric about the
mid-plane, the same poloidal structure, and the same field strength in the inner disc. The
evolution of the protostar mass, angular momentum and magnetic energy is likewise
indistinguishable from that of the 1MsunHLL run without enhanced resistivity over the interval
covered by the test.

We conclude that the inner boundary produces a genuine but strictly local artefact, restricted
to a few cells, and that the transport results reported in Sects.~\ref{sec: angular_momentum_equilibrium} and~\ref{sec: transport_mechanisms} are not
affected by it.

\end{appendix}

\end{document}